\documentclass[fleqn,usenatbib,usedcolumn]{mnras}

\usepackage{mathptmx}

\usepackage[T1]{fontenc}
\usepackage{ae,aecompl}

\usepackage{graphicx}	
\usepackage{amsmath}	
\usepackage{amssymb}	
\usepackage{breqn}
\usepackage{float}
\usepackage{placeins}

\title[30 - 40 $\mu$m AGN Observations on SOFIA]{SOFIA/FORCAST Resolves 30 - 40 $\mu$m Extended Dust Emission in Nearby Active Galactic Nuclei}

\author[L. Fuller et al.]{
Lindsay Fuller,$^{1}$\thanks{E-mail: lindsay.fuller@utsa.edu}, Enrique Lopez-Rodriguez$^{2}$, Chris Packham,$^{1,3}$, Kohei Ichikawa$^{1,3,4,5,6}$,
\newauthor   Aditya Togi$^{1,7}$, Almudena Alonso-Herrero$^{8,1}$, Cristina Ramos-Almeida$^{9,10}$,
\newauthor  Tanio Diaz-Santos$^{11}$, N. A. Levenson$^{12}$, James Radomski$^{2}$
\\
$^{1}$The University of Texas at San Antonio, One UTSA Circle, San Antonio TX, 78249, USA\\
$^{2}$SOFIA Science Center, NASA Ames Research Center, Moffett Field, CA 94035, USA\\
$^{3}$National Astronomical Observatory of Japan, 2-21-1 Osawa, Mitaka, Tokyo 181-8588, Japan \\
$^{4}$Department of Astronomy, Columbia University, 550 West 120th St, New York, NY 10027, USA \\
$^{5}$Frontier Research Institute for Interdisciplinary Sciences, Tohoku University, Sendai, Miyagi 980-8578, Japan\\
$^{6}$Astronomical Institute, Tohoku University, Aramaki, Aoba-ku, Senda, Miyagi 980-8578, Japan\\
$^{7}$St. Mary's University, 1 Camino Santa Maria, San Antonio, TX, 78228, USA\\
$^{8}$Centro de Astrobiolog\'{\i}a (CSIC-INTA), ESAC Campus, E-28692 Villanueva de la Ca\~nada, Madrid, Spain\\
$^{9}$Instituto de Astrof\' isica de Canarias, Calle V\' ia L\'actea, s/n, E-38205, La Laguna, Tenerife, Spain\\
$^{10}$Departamento de Astrof\' isica, Universidad de La Laguna, E-38206, Tenerife, Spain\\
$^{11}$Núcleo de Astronomía de la Facultad de Ingeniería, Universidad Diego Portales, Av. Ejercito Libertador 441, Santiago, Chile\\
$^{12}$Space Telescope Science Institute, 3700 San Martin Drive, Baltimore, MD, 21218, USA\\
}

\date{Accepted XXX. Received YYY; in original form ZZZ}

\pubyear{2018}

\begin{document}
\label{firstpage}
\pagerange{\pageref{firstpage}--\pageref{lastpage}}
\maketitle

\begin{abstract}
We present arcsecond-scale observations of the active galactic nuclei (AGNs) of seven nearby Seyfert galaxies observed from the Stratospheric Observatory For Infrared Astronomy (SOFIA) using the 31.5 and 37.1 $\mu$m filters of the Faint Object infraRed CAmera for the SOFIA Telescope (FORCAST).  We isolate unresolved emission from the torus and find extended diffuse emission in six 37.1 $\mu$m residual images in our sample.  Using $Spitzer$/IRS spectra, we determine the dominant mid-infrared (MIR) extended emission source and attribute it to dust in the narrow line region (NLR) or star formation.  We compare the optical NLR and radio jet axes to the extended 37.1 $\mu$m emission and find coincident axes for three sources. For those AGNs with extended emission coincident with the optical axis, we find that spatial scales of the residual images are consistent with 0.1 - 1 kpc scale distances to which dust can be heated by the AGN.  Using previously published subarcsecond 1 - 20 $\mu$m imaging and spectroscopic data along with our new observations, we construct broadband spectral energy distributions (SEDs) of the AGNs at wavelengths 1 - 40 $\mu$m. We find that three AGNs in our sample tentatively show a turnover in the SED between 30 - 40 $\mu$m.  Using results from \textsc{Clumpy} torus models and the Bayesian inference tool \textsc{BayesClumpy}, we find that the posterior outputs for AGNs with MIR turnover revealed by SOFIA/FORCAST have smaller uncertainties than AGNs that do not show a turnover.
\end{abstract}




\section{Introduction}

\defcitealias{F16}{F16}

Infrared (IR) emission from active galactic nuclei (AGNs) is largely attributed to an optically and geometrically thick toroidal dust structure that primarily intercepts optical and ultraviolet (UV) emission from the central accreting black hole and re-emits in IR wavelengths.  According to the unified model \citep{Antonucci1993,UP1995}, an approximate edge-on view of the thick dust torus fully obscures the center in Type 2 AGNs, while a more pole-on view allows a direct line of sight into the center of Type 1 AGNs.  See \citet{Netzer2015,Ramos2017} for recent reviews on nuclear obscuration.  

The AGN torus has been modeled assuming a smooth dust distribution \citep{Pier1992,Pier1993,Granato1994,Siebenmorgen2004,Schartmann2005} and, more recently, a clumpy distribution \citep{Nenkova2008a,Honig2010}.  The latter more accurately describes imaging \citep{Packham2005,Radomski2008}, interferometry \citep{Jaffe2004,Tristram2007,Raban2009,Burtscher2013,Tristram2014}, and recent ALMA observations of NGC 1068 \citep{GB2016}, which establish a parsec-scale torus.  The clumpy models of \citet{Nenkova2008a} have been used extensively to derive torus model parameters \citep{RA2009,RA2011,AH2011,Ichikawa2015,MP2017}.  The radial size of the torus gives insight as to the wavelength of peak IR emission \citep{vBD2003,RA2011,AR2013}.  However, due to a lack of high spatial resolution (< 1") observations beyond 20 $\mu$m, there has not yet been observational confirmation of the wavelength of peak flux density of the torus.  Models suggest that the turnover of the torus emission is between 20 - 50 $\mu$m  \citep{Nenkova2008a,Honig2010,RA2014}.  Using data from the 2.5-m Stratospheric Observatory For Infrared Astronomy (SOFIA), \citet{F16} (hereafter \citetalias{F16}) found that the wavelength of peak flux density generally does not occur at wavelengths less than 31.5 $\mu$m. 

The SOFIA telescope provides the best spatial resolution ($\sim$ 3.4" at 31 $\mu$m) to date for observations between $\sim$ 30 - 70 $\mu$m.  However, SOFIA cannot resolve the subarcsecond-scale torus structure and contamination from other MIR sources is likely.  Spectroscopic analysis of polycyclic aromatic hydrocarbon (PAH) emission features reveals that excess MIR emission in some AGNs can have a major contribution from star formation \citep[SF;][]{Clavel2000,Weedman2005,Buchanan2006,Tommasin2008,Ichikawa2012,Esquej2014}.  For example, \citet{AH2014} found that PAH emission in six AGNs known to have nuclear star formation occurs at distances of $\sim$ 60 - 420 pc.  

There is also significant evidence for IR emission coincident with the narrow line region (NLR) in some AGNs.  On subarcsecond scales, \citet{Honig2012,Honig2013,Tristram2014,LG2016} detected MIR emission distributed in the polar regions of the nucleus of several Seyfert galaxies. On larger scales, \citet{Asmus2016} detected clear extended polar emission consistent with the NLR out to hundreds of parsecs for 18 AGN.  \citet{Mason2006} found that MIR emission of NGC 1068 emanates from two separate components; the torus dominates emission < 1", but for apertures $\gtrsim$ 1", MIR emission is dominated by dust emission from the ionization cones.  Likewise, \citet{Radomski2003} resolves 10.8 and 18.2 $\mu$m emission in NGC 4151 extending 3.5" in the direction of the NLR.

In this paper, we present the best spatial resolution 37.1 $\mu$m imaging data currently available for a sample of 7 bright Seyfert galaxies using the SOFIA telescope.  Of those 7, we also obtained 31.5 $\mu$m imaging data for 3 AGN.  We examine the turnover of emission from the obscuring torus, and also determine the source of extended emission for each AGN in our sample. To provide a description of the physical properties of the torus, we use the \textsc{Clumpy} torus models of \citet{Nenkova2008a} with the Bayesian inference tool, \textsc{BayesClumpy} \citep{AR2009}, to fit the IR (1 - 37.1 $\mu$m) nuclear SEDs.  In Section \ref{observations_section} we describe the observations and data reduction; in Section \ref{imaging} we explain our method of image analysis;  in Section \ref{discussion} we discuss the source of extended emission in our sample;  in Section \ref{bayes} we describe the model fitting and results; and Section \ref{conclusions} summarizes our analysis.


\section{Observational Data}
\label{observations_section}

\subsection{Sample Selection}

The primary motivation of this study is to extend the wavelength range of AGN observations to 37.1 $\mu$m for use in modeling MIR emission around the active nucleus.  Three sources (NGC 3081, NGC 3227, and NGC4388) were included in the study of \citetalias{F16} in the wavelength range between 1 - 31.5 $\mu$m.  One source (NGC 4151) was not included in \citetalias{F16}, but has been extensively studied from 1 - 18 $\mu$m \citep{RA2009,AH2011,Ichikawa2015}.  Three sources (Mrk 3, NGC 1275, and NGC 2273) were chosen based on their inclusion in the N-band spectral atlas of \citet{AH2016}.

The seven Seyfert galaxies presented in this work are part of an on-going AGN survey within the 30 - 100 $\mu$m range using SOFIA (Proposal ID: \#02\_0035, \#04\_0048, \#06\_0066 PI: Lopez-Rodriguez).  The AGN survey includes bright, nearby Seyfert 1 and 2 galaxies in a flux-limited sample with nuclear fluxes > 500 mJy at 31.5 $\mu$m, and bolometric luminosities $43 \le \log$ $L_{\rm bol} ({\rm erg}~{\rm s}^{-1}$) $\le 46$.  In the current sample, we selected galaxies with subarcsecond-resolution N-band spectroscopy from the literature.  The sample is not representative, and the high-resolution spectra are included in order to represent only emission from the torus surrounding the AGN.  The basic properties of the sample are given in Table \ref{info}.  


\begin{table*}
\begin{minipage}{100mm}
\caption{AGN properties}
\centering
\begin{tabular}{ccccccc}
\hline
\hline
Object		&	Type		& 	$z$	&	Distance		&	Scale 	&	log $L_{\rm bol}$	& Ref.		\\
			&			&		&	(Mpc)		& (pc/")		&	(erg s$^{-1}$)		&			\\
\hline
Mrk 3		&	Sy2		&	0.0135	&	54.9		&	266		&  45.1 & 1,i,a			\\
NGC 1275	&	RG/Sy1.5	&	0.0176	&	71.5		&	346		&  44.8 & 2,ii,b				\\
NGC 2273	&	Sy2		&	0.0061	&	25.0		&	121		&  43.9   & 3,iii,c			\\
NGC 3081	&	Sy2		&	0.0080	&	32.4		&	157		&  44.2  & 4,iv,a					\\
NGC 3227	&	Sy1.5	&	0.0039	&	15.7		& 	76		&  43.3   & 5,v,a				\\
NGC 4151	&	Sy1.5	&	0.0033	&	13.0		&	63		&  43.9  & 5,v,d				\\
NGC 4388	&	Sy2		&	0.0047	&	19.0		&	92		&  44.7 & 6,vi,a  					\\
\hline
\hline
\end{tabular}\\
\textsc{References:}  Seyfert type: 1) \citet{Khackikian1974}, 2) \citet{AH2016}, 3) \citet{Contini1998}, 4) \citet{Phillips1983}, 5)  \citet{Veron2010}, 6) \citet{Trippe2010}.
Redshift: i) \citet{Tifft1988}, ii) \citet{Strauss1992}, iii) \citet{Ruiz2005}, iv) \citet{Theureau2005},  v) \citet{deVaucouleurs1991} vi) \citet{Lu1993}, Distances were calculated using $H_{0}$ = 73.8 km s$^{-1}$Mpc$^{-1}$.
Luminosities: a) \citet{Ichikawa2017} b)\citet{Baumgartner2013}, c) \citet{Marinucci2012}, d) \citet{Marconi2004}
\label{info}
\end{minipage}
\end{table*}

\subsection{Observations}

We obtained photometric observations (Proposal ID: \#04\_0048, PI: Lopez-Rodriguez) using the 31.5 and 37.1 $\mu$m filters of the Faint Object infRared CAmera for the SOFIA Telescope \citep[FORCAST;][]{Herter2012} on the SOFIA airborne observatory \citep{Young2012}.  FORCAST is a dual-channel camera and spectrograph, operational from 5 to 40 $\mu$m.  The short wavelength channel (SWC; $\lambda$ < 25 $\mu$m) and the long wavelength channel (LWC; $\lambda$ > 25 $\mu$m) have a 256 $\times$ 256 pixel$^{2}$ array and can be used in simultaneous dual channel mode or individually.  The 0.768 "/pixel scale gives an effective field of view (FOV)  of 3.4 $\times$ 3.2 arcmin$^{2}$.  Our data was taken with the 31.5 $\mu$m ($\Delta\lambda$ = 5.7 $\mu$m) and the 37.1 $\mu$m ($\Delta\lambda$ = 3.3 $\mu$m) filter used with the LWC in single channel mode.

Observations were made using the two-position chop-nod (C2N) method with symmetric nod-match-chop (NMC) to remove time-variable sky background and telescope thermal emission, and to reduce the effect of 1/$f$ noise from the array.  Data were reduced using the \textsc{forcast\_redux} pipeline version 1.1.3 as described in \citet{Herter2012}.  NGC 1275 was observed twice during the cycle, once in February 2016 and once in September 2016.  We present the results as an average of the two observations.  A summary of observations, including the FWHM of the AGN and point spread functions (PSFs), are given in Table \ref{observations}.  The variation in the 31.5 $\mu$m standard FWHM was not found to be wavelength dependent, but rather due to variability in seeing or tracking on SOFIA. 

\subsection{Point Spread Functions} 
\label{psf}

\begin{figure*}
\includegraphics[scale=0.75]{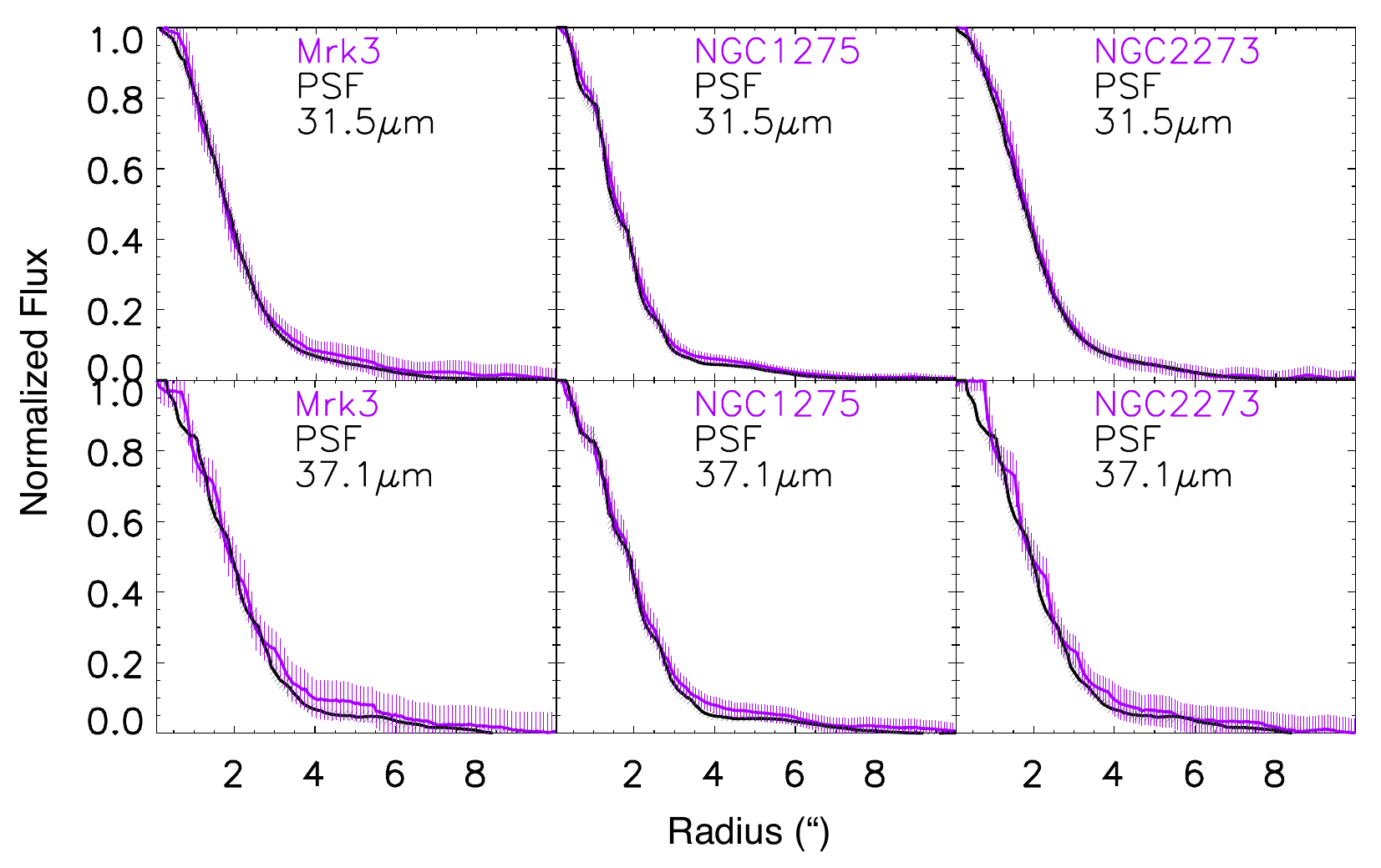}
\includegraphics[scale=0.8]{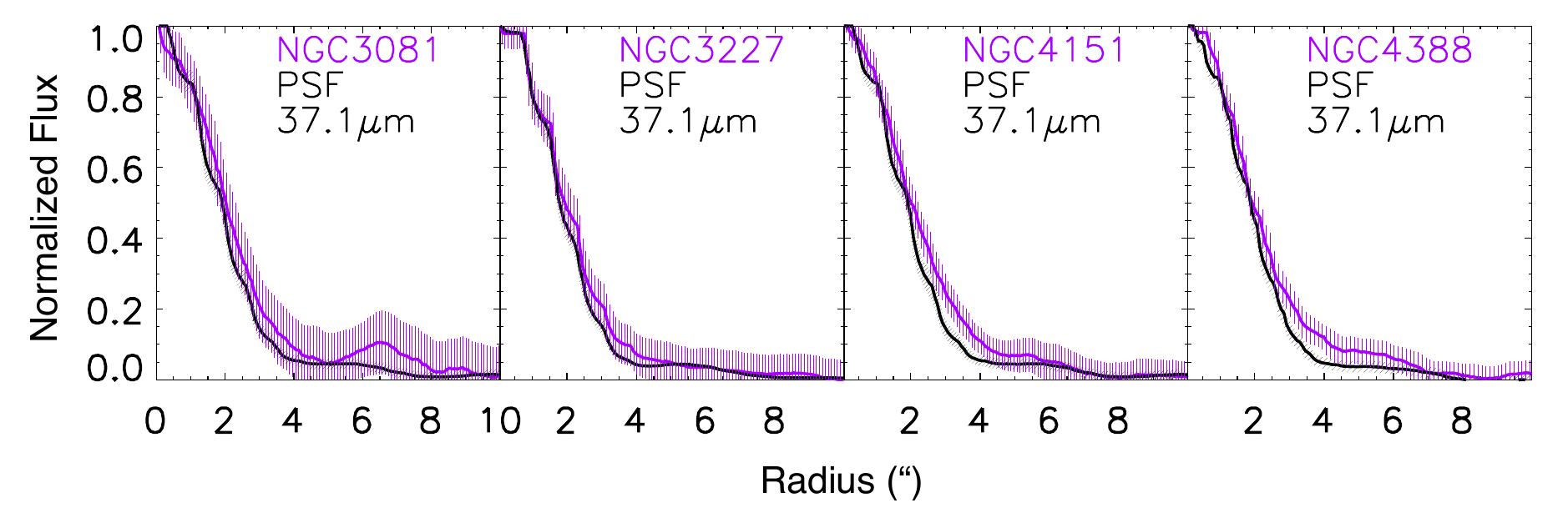}
\caption{Radial profiles of the PSFs (solid black line) constructed as described in Section \ref{psf} compared to those of the AGNs (solid violet line).  The regions shaded with horizontal bars indicate uncertainty from the background and from variations in the signal at increasing radii from the center.  The top panel contains the 3 AGNs observed in both the 31.5 and 37.1 $\mu$m filters, whereas the bottom panel contains the 4 AGNs observed in only the 37.1 $\mu$m filter. } 
\label{profiles}
\end{figure*}

The PSF was taken to be the average of multiple observations of a single standard star for a given observing date in the corresponding filter.  On 17 February 2016 two separate standards were observed at 37.1 $\mu$m.  For that data, the standard star observed in the time frame closest to the AGN observation was the averaged standard used for the PSF.  The radial profiles of the AGN observations and their associated PSFs are shown in Figure \ref{profiles}.


\begin{table*}
\begin{minipage}{140mm}
\caption{Observational data}
\centering
\begin{tabular}{cc|cc|cc|c|c}
\hline
\hline
Object	& 	Standard	&        \multicolumn{2}{c}{PSF FWHMs}		&	 \multicolumn{2}{c}{AGN FWHMs} 		&	Observing Date	&	t$_{on-source}$ 	\\
\hline
		&	&31.5 $\mu$m			&	37.1 $\mu$m	&	31.5 $\mu$m	& 37.1 $\mu$m 		&  	&			\\
		&	&(arcsec$^{2}$)				&	(arcsec$^{2}$)		&	(arcsec$^{2}$)		& (arcsec$^{2}$)			& (yyyy-mm-dd)	& (seconds)	\\
\hline
Mrk 3		&	$\beta$ UMi	&	4.57 $\times$ 4.32	&	4.62 $\times$ 4.63	&	4.58 $\times$ 4.13	&	5.28 $\times$ 4.58	&	2016-09-27		&		288, 	390		 	\\
NGC 1275	&	$\gamma$ Dra	&	4.05 $\times$ 3.68	&	4.61 $\times$ 4.34	&	3.85 $\times$ 4.31	&	4.87 $\times$ 4.55	&	2016-02-06		&		292, 	397	\\
			&	$\beta$ And	&	4.13 $\times$ 3.77	&	4.56 $\times$ 4.28	&		\dots			&		\dots			&	2016-09-21		&		288, 339	\\
NGC 2273	&	$\beta$ UMi	&	4.57 $\times$ 4.32	&	4.62 $\times$ 4.63	&	4.65 $\times$ 4.24	&	5.13 $\times$ 4.78	&	2016-09-27		&		547, 	700	\\
NGC 3081	&	$\alpha$ Boo	&		\dots			&	4.59 $\times$ 4.35	&		\dots			&	4.70 $\times$ 4.81	&	2016-02-17		&		699\\
NGC 3227	&	Europa		&		\dots			&	4.47	$\times$ 4.19	&		\dots			&	4.92 $\times$ 4.66	&	2016-02-17		&		705	\\
NGC 4151	&	$\alpha$ Boo	&		\dots			&	4.59 $\times$ 4.35	&		\dots			&	5.01 $\times$ 5.45	&	2016-02-17		&		369	\\
NGC 4388	&	$\gamma$ Dra	&		\dots			&	4.61 $\times$ 4.34	&		\dots			&	5.67 $\times$ 4.50	&	2016-02-06		&		378\\
\hline	
\hline
\end{tabular}
Column 2: Name of the PSF standard star. Columns 3,4: Major and minor axes of the 31.5, 37.1 $\mu$m PSF FWHM. Columns 5,6: Major and minor axes of the 31.5, 37.1 $\mu$m AGN FWHM. Column 7: Date of observations. Column 8: On source time for the 31.5 and 37.1 $\mu$m filters.  Some PSF standards are used for multiple science targets.
\label{observations}
\end{minipage}
\end{table*}

\section{Imaging Analysis}
\label{imaging}

\begin{figure*}

\includegraphics[scale=0.75]{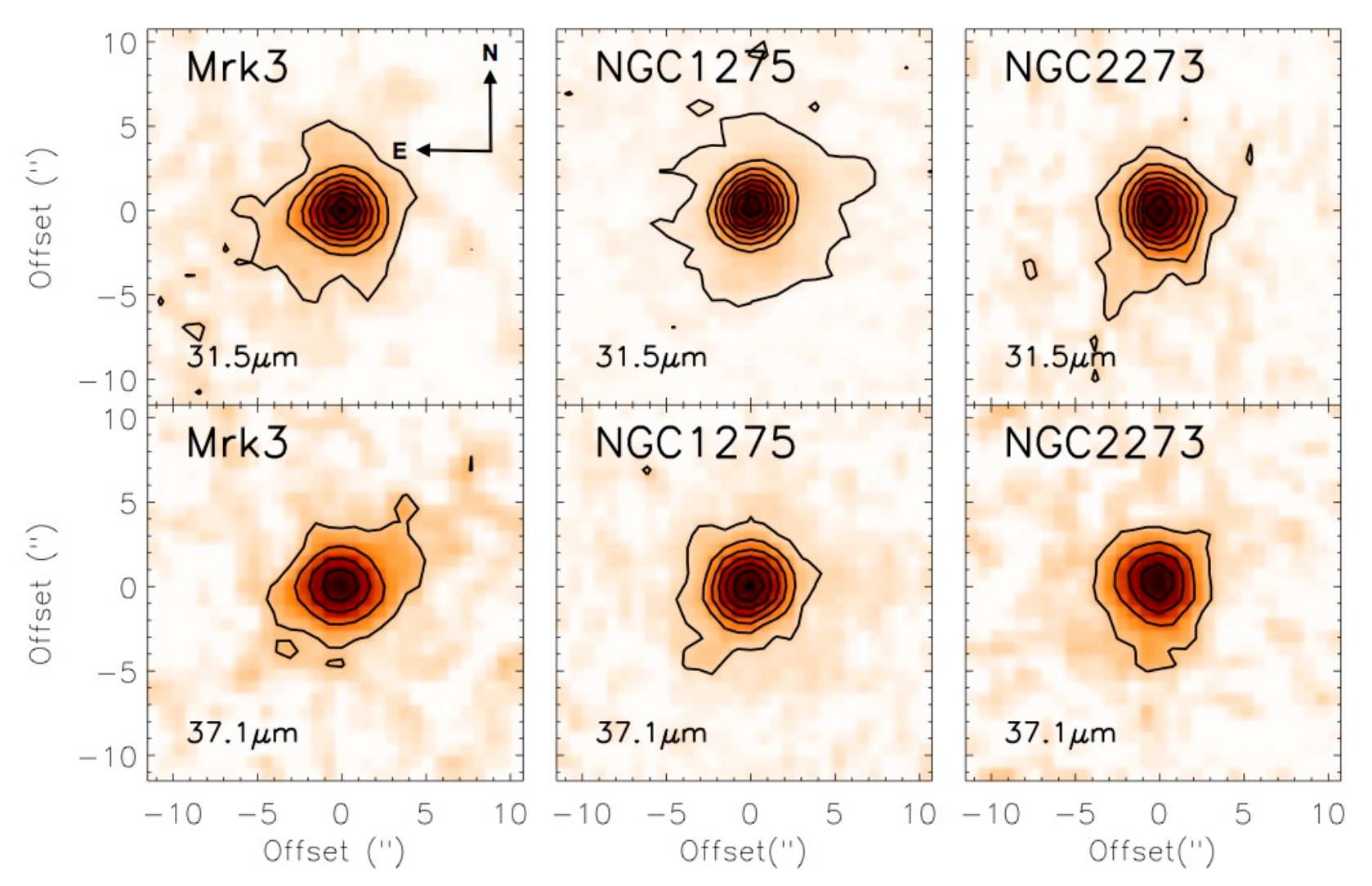}
\includegraphics[scale=0.75]{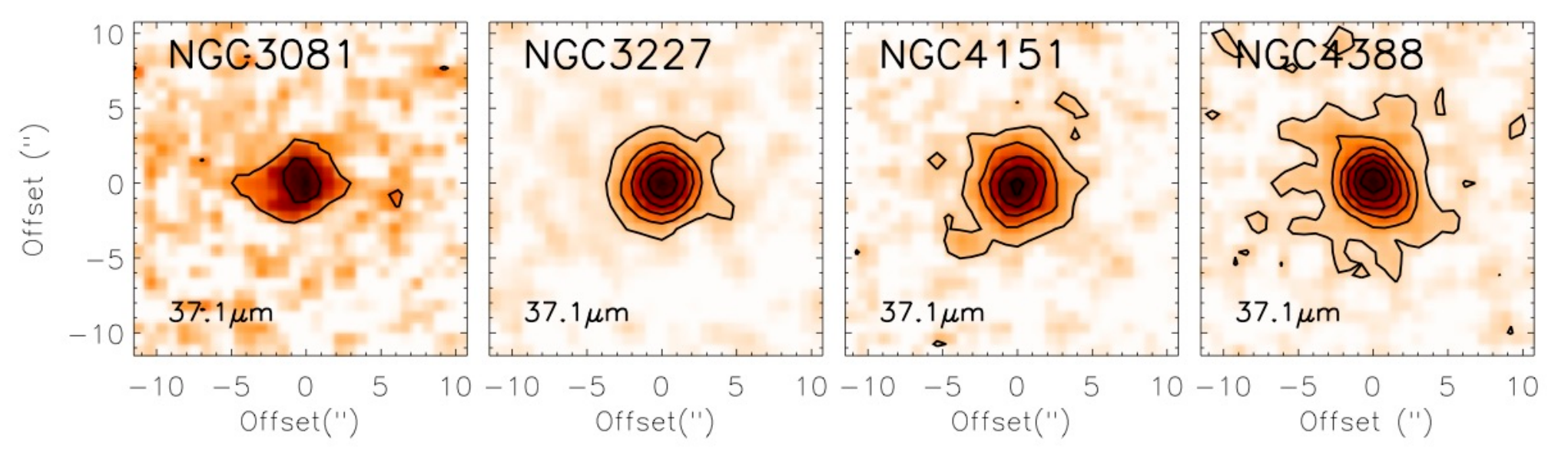}
\caption{SOFIA/FORCAST 31.5 and 37.1 $\mu$m filter images of AGN sample.  Each mosaic is a 20" $\times$ 20" image centered on the position of peak emission.  North is up, east is to the left.  Top: Galaxies observed at both 31.5 and 37.1 $\mu$m.  Bottom:  Galaxies observed at only 37.1 $\mu$m.  Lowest contours are 3$\sigma$ and increase in steps of 5$\sigma$ (contours of NGC 1275 increase in steps of 10$\sigma$).}
\label{images}
\end{figure*}

Figure \ref{images} shows a 20" $\times$ 20" FOV of the seven newly-observed AGNs in our sample, centered on the peak of each source.  In Table \ref{observations} we show a comparison of the FWHM of the PSFs to the FWHM of the AGNs.  All objects show a larger FWHM than the PSF FWHM at 37.1 $\mu$m.  In Section \ref{discussion} we examine the source of extended emission.

To account for host galaxy contamination in the nuclear fluxes, we used the PSF scaling method described in \citetalias{F16}.  The total flux density is measured in an 18" diameter circular aperture (an 18" aperture is used in the flux calibration to ensure as close to 100\% of the standard star flux is measured).  The PSF of the observation, scaled to the peak of galaxy emission, represents the maximum contribution from the unresolved torus component.  The scale of the PSF is reduced until the subtraction of the PSF from the source image (hereafter called the \textit{residual}) yields a smooth profile.  Due to MIR sky variation on the order of $\sim$10\%, we use a 10\% increment in scaling the PSF as in previous works \citep[e.g.][]{Radomski2002,Radomski2003,Packham2005,Levenson2009,RA2009,RA2011,GB2015,F16}. Table \ref{flux} gives the total flux within an 18" aperture ($F_{\rm tot}$), the total percent the PSF was scaled (\% $PSF_{\rm scale}$), and the PSF-scaled flux ($F_{\rm PSF}$) which represents the flux from the unresolved torus component.  The described method yielded inconsistent results for NGC 3081.  Due to this and the low signal-to-noise ratio (SNR), we use $F_{\rm tot}$ as an upper limit.  This is discussed further in Section \ref{ngc3081}.

To estimate photometric uncertainty, we first determined the variability in individual calibration factors for the standard stars associated with a given observation date, which was $\sim$6\%.  We determined that the uncertainty due to a variable PSF obtained by cross-calibrating the standard stars is $\sim$7\%.  Additional uncertainty in unresolved fluxes determined from the PSF subtraction was estimated as 10\%.  The total uncertainties were estimated by adding these contributions in quadrature.

\begin{table*}

\begin{minipage}{100mm}
\caption{31.5 and 37.1 $\mu$m flux densities}
\begin{tabular}{c|ccc|ccc}
\hline
\hline
Object		&			&	31.5 $\mu$m		&				&			&		37.1 $\mu$m			&			\\
\hline

			&	$F_{\rm tot}$ 		&	$PSF_{\rm scale}$ & $F_{\rm PSF}$	&	$F_{\rm tot}$ 		&	$PSF_{\rm scale}$ &	 $F_{\rm PSF}$	\\
			&		(Jy)			&	(\%)				&	 (Jy)	&	(Jy)				&	(\%)				& (Jy)			\\
\hline
Mrk 3		&	2.9   $\pm$ 0.3		&	70		&	1.8   $\pm$ 0.2 		&	3.0   $\pm$ 0.3 		&	70	&	1.9   $\pm$ 0.3		\\
NGC 1275	&	4.0   $\pm$ 0.5		&	80		&	3.0	$\pm$ 0.5		&	5.0   $\pm$ 0.8		&	70	&	3.2	$\pm$ 0.7 	\\
NGC 2273	&	1.9   $\pm$ 0.2		&	55		&	1.0   $\pm$ 0.1		&	2.7   $\pm$ 0.3 		&	50	&	1.2   $\pm$ 0.1		\\
NGC 3081	&	0.8   $\pm$ 0.1		&	100		&	0.8   $\pm$ 0.1  	&	1.4   $\pm$ 0.2		&	100	&	1.4   $\pm$ 0.2		\\
NGC 3227	&		\dots			&	\dots		&	\dots				&	2.8   $\pm$ 0.3		&	45	&	1.1   $\pm$ 0.1		\\
NGC 4151	&		\dots			&	\dots		&	\dots				&	4.5   $\pm$ 0.6		&	70	&	2.7   $\pm$ 0.5		\\
NGC 4388	&		\dots			&	\dots		&	\dots				&	3.2   $\pm$ 0.3		&	70	&	1.8   $\pm$ 0.3		\\
\hline
\hline
\end{tabular}
Column 2: Total flux of the AGN observation; Column 3: \% of the PSF scaling; Column 4: Flux of unresolved torus; Columns 5-7: same as columns 2-4 for 37.1 $\mu$m observations.
\label{flux}
\end{minipage}
\end{table*}

\begin{figure}
\centering
\includegraphics[scale=0.7]{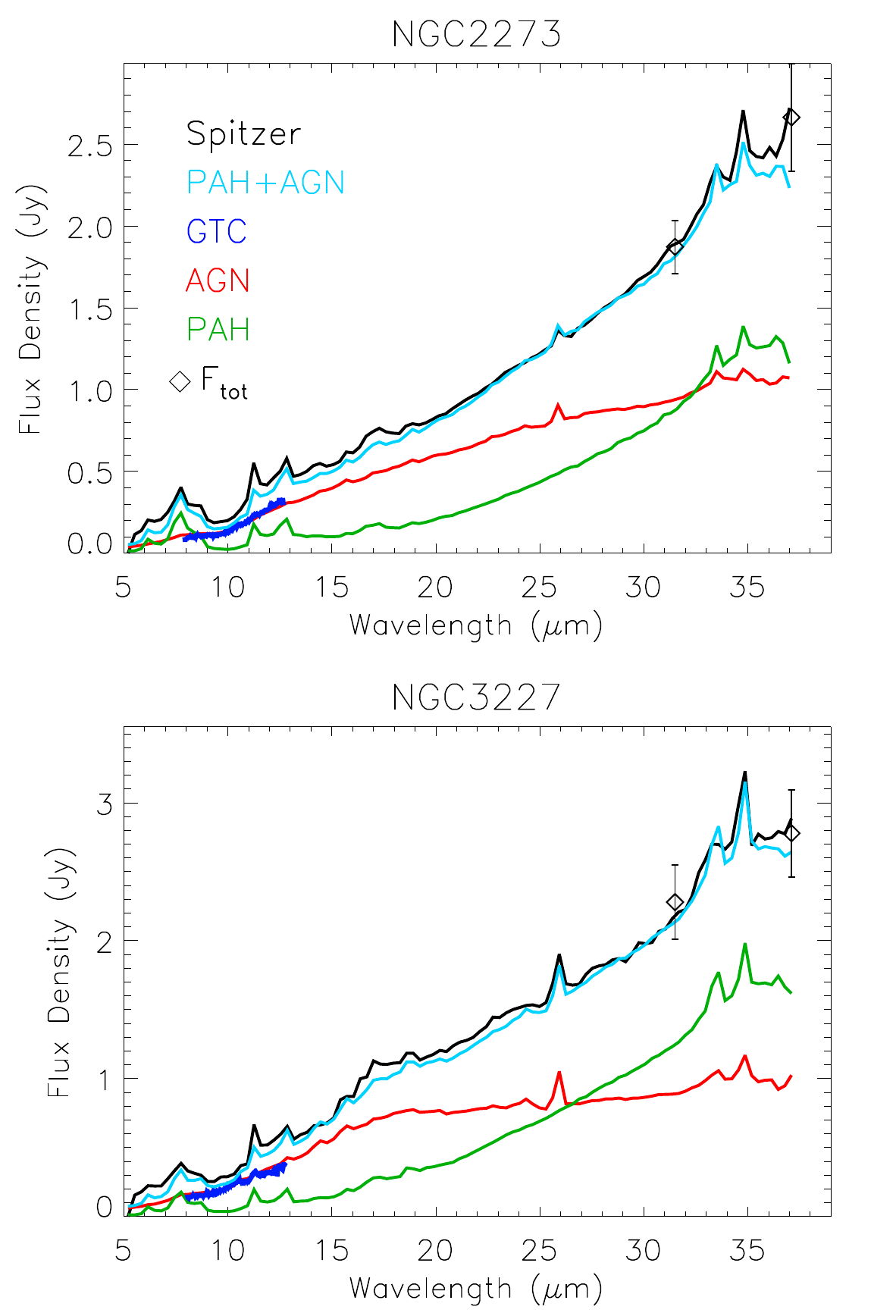}
\caption{A spectral decomposition with a reduced $\chi^{2}$ was only available for two objects: NGC 2273 and NGC 3227.  Results of the decomposition using DeblendIRS \citep{HC2015} are shown here.  Spectra are separated into a PAH component (green) and the AGN component (red).  The sum of the two components is shown here in light blue, which coincides with the original $Spitzer$ spectra in black.  The AGN component coincides with with sub-arcsecond GTC spectra (dark blue), while the total flux from SOFIA coincides with the $Spitzer$ continuum.}
\label{deblendirs}
\end{figure}

Following the method of \citetalias{F16} to test the results of the PSF scaling, we performed a spectral decomposition analysis of $Spitzer$/IRS spectra.  The routine DeblendIRS \citep{HC2015} uses a combination of three templates to represent total MIR emission from $Spitzer$/IRS; these emission sources are 1) AGN, 2) star formation (PAH), and 3) diffuse host galaxy emission.  See also \citet{GB2016} for an example of DeblendIRS applied to $Spitzer$ spectra of Seyfert galaxies.  The routine has templates for each component and outputs the combination of templates with the lowest $\chi^{2}$.  The $Spitzer$ spectra generally span 5 - 38 $\mu$m, however many of the templates are from high redshift sources, so their spectra reach 38 $\mu$m in the observed frame, but not in the rest frame.  For this reason only 20 of 189 AGN templates were applicable to observations at 37.1 $\mu$m.  Due to the reduced number of templates, the decomposition showed a reduced $\chi^{2}$ ($\sim$2) for only two AGNs, NGC 2273 and NGC 3227.  The PSF scaling results for these two AGNs were not consistent with the spectral decomposition.  For this reason, we scaled the PSF to match the AGN contribution from the decomposition and used the results as upper limits.  Figure \ref{deblendirs} shows the results of the decomposition, high-resolution spectra from the GTC, and the total flux from SOFIA.

\section{Origin of Extended Emission}
\label{discussion}

Our PSF-scaling results show residual extended galaxy emission at 37.1 $\mu$m for the six objects in our sample on which we performed the scaling.  In this section we examine the origin of the extended residual emission, which would ostensibly originate from star forming regions, the NLR, or possibly even the torus or host galaxy.  We use only the extended 37.1 $\mu$m emission since it is the common filter between each AGN in this sample.

MIR emission is substantially less affected by extinction from the host galaxy than optical or UV observations.  PAHs, which are indicative of stellar processes, emit strongly in MIR spectra \citep{Roche1991,Genzel1998,Sturm2000,FS2004,Peeters2004}, the most prominent and recognizable emission line features occurring at 3.3, 6.2, 7.7, 8.6, 11.3, 12.7, and 17.0 $\mu$m.  PAHs absorb mostly UV photons \citep{Uchida1998} and are associated with photon energies between 6 - 13.6 eV.  

\citet{Genzel1998} show a direct correlation between AGN activity and the strength of high ionization lines such as [Ne \textsc{v}] 14.3, 24.3 $\mu$m and [O \textsc{iv}] 25.9 $\mu$m because of their high ionization potentials.  [Ne \textsc{v}] 14.3, 24.3 $\mu$m, is commonly used to distinguish AGN related activity \citep{Sturm2000,Abel2008} from the NLR and not star formation since the production of Ne$^{4+}$ requires photons with energies greater than $\sim$ 97 eV.  Likewise, [O \textsc{iv}] 25.9 $\mu$m is an effective tracer of AGN activity \citep{Melendez2008,Rigby2009,DS2009} with an ionization potential of $\sim$ 55 eV.

\citet{Asmus2016} use [\textsc{O iv}] 25.9 $\mu$m as a tracer of NLR activity and find a direct correlation between strong [\textsc{O iv}] emission and nuclear extension.  Eighteen objects in their sample showed extended MIR emission aligned with the the [O \textsc{iii}] $\lambda$5007 and radio axes.  \citet{Haniff1988} found that, in a sample of 10 Seyfert galaxies, [O \textsc{iii}] emission is aligned with the radio axis to within a few degrees.  By comparing $HST$ optical observations to radio observations, \citet{Falcke1998} found that radio jets interact with gas near the AGN and can affect the morphology of the NLR.  Hence, [O \textsc{iv}] can probe the NLR, which should be aligned with the optical and radio axes.  

In the following subsections, we use the 37.1 $\mu$m residual images from the PSF-scaling as well as redshift-corrected 5 - 38 $\mu$m spectra from the $Spitzer$ CASSIS library \citep{Lebouteiller2011} to determine the origin of extended emission on arcsecond scales.  We compare the residual image with optical observations, which were smoothed to reduce background noise, as well as the radio axis P.A.  We use PAH features at 6.2, 7.7, 8.6, and 11.3 $\mu$m as diagnostics for star formation, but do not use the 12.7 or 17.0 $\mu$m lines due to blending with H$_{2}$ and [Ne \textsc{ii}] 12.8 $\mu$m.  We also compare the arcsecond-scale spectra from $Spitzer$ to subarcsecond N-band spectra (Table \ref{spectroscopy}) and determine the amount of extended 12 $\mu$m emission.  This wavelength was chosen because it is not heavily affected by 9.7 $\mu$m silicate absorption and also avoids the main 11.3 and 12.7 $\mu$m PAH features, though the amount of extended emission is highly variable in this wavelength range.

\begin{table}

\begin{minipage}{80mm}
\centering
\caption{Sub-arcsecond spectroscopy}

\begin{tabular}{cccc}
\hline
Object	&	Instrument	&	Slit Width (")	&	Ref.		\\
\hline
Mrk 3	&	GTC/CanariCam			&	0.52	&	a	\\
NGC 1275	&	GTC/CanariCam		&	0.52	&	a	\\
NGC 2273	&	GTC/CanariCam		&	0.52	&	a	\\
NGC 3081	&	Gemini/T-ReCS			&	0.65	&	b	\\
NGC 3227	&	GTC/CanariCam		&	0.52	&	a	\\
NGC 4151	&	Gemini/Michelle		&	0.36	&	c	\\
NGC 4388	&	GTC/CanariCam		&	0.52	&	a	\\
\hline

\end{tabular}
\label{spectroscopy}
\end{minipage}

\textsc{References:}, a) \citet{AH2016}, b) \citet{GM2013}, c) \citet{AH2011}

\end{table}

\begin{figure*}

\includegraphics[scale=0.38]{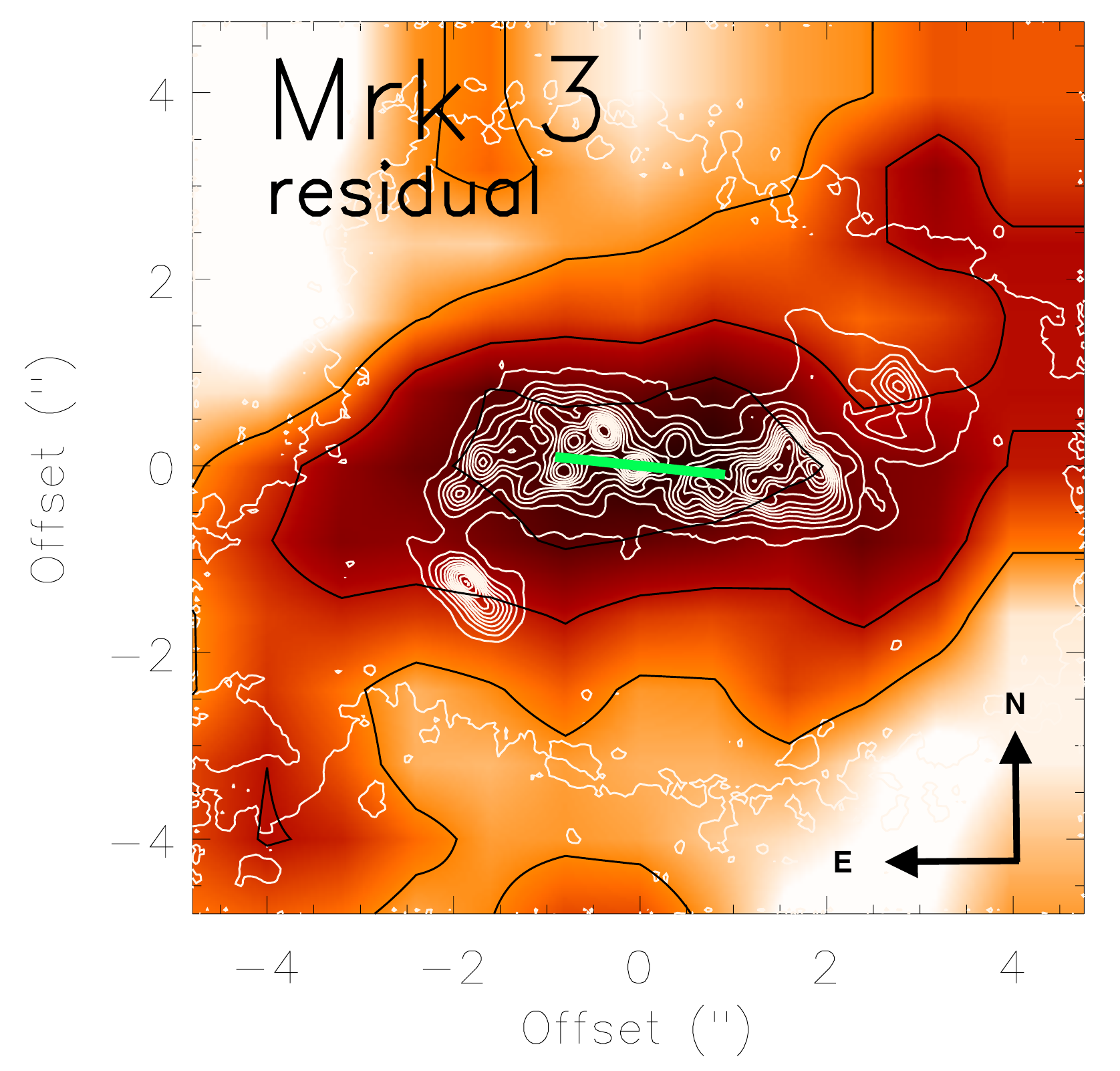}
\includegraphics[scale=0.74]{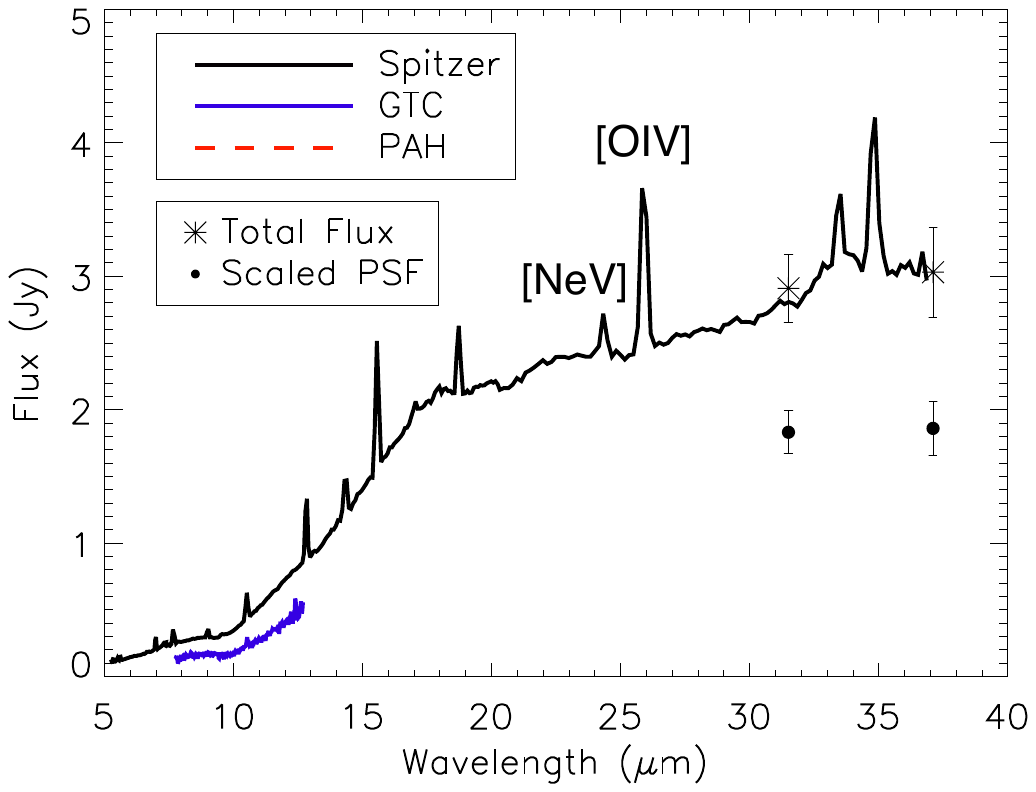}
\includegraphics[scale=0.38]{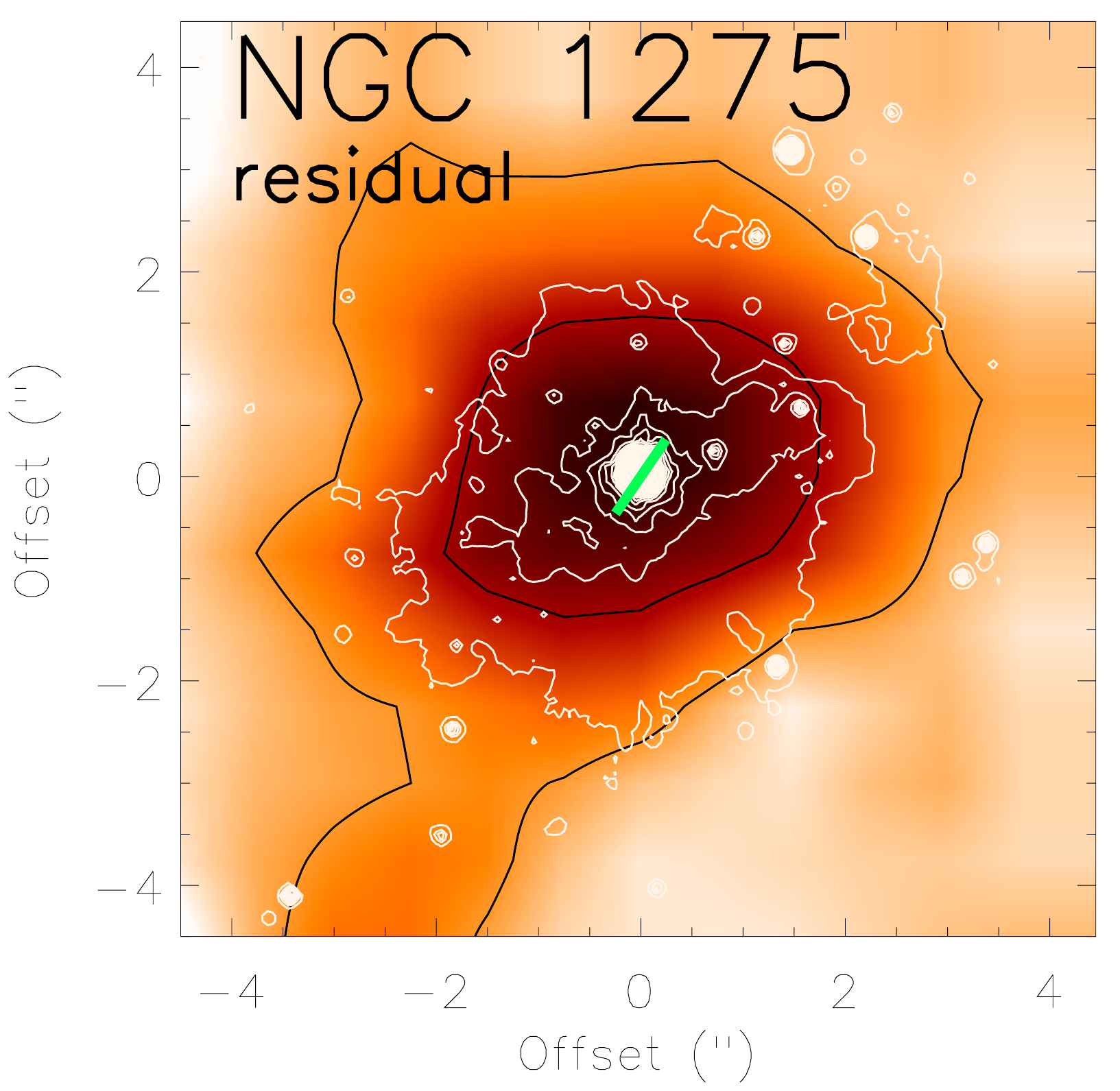}
\includegraphics[scale=0.75]{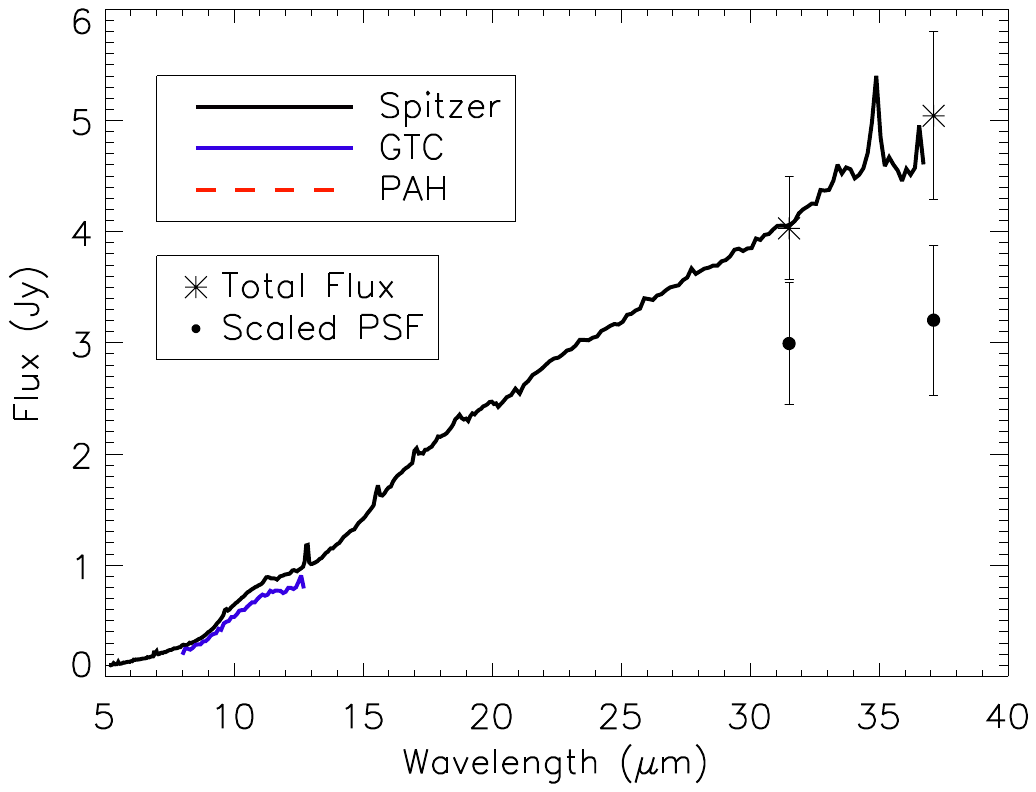}
\includegraphics[scale=0.385]{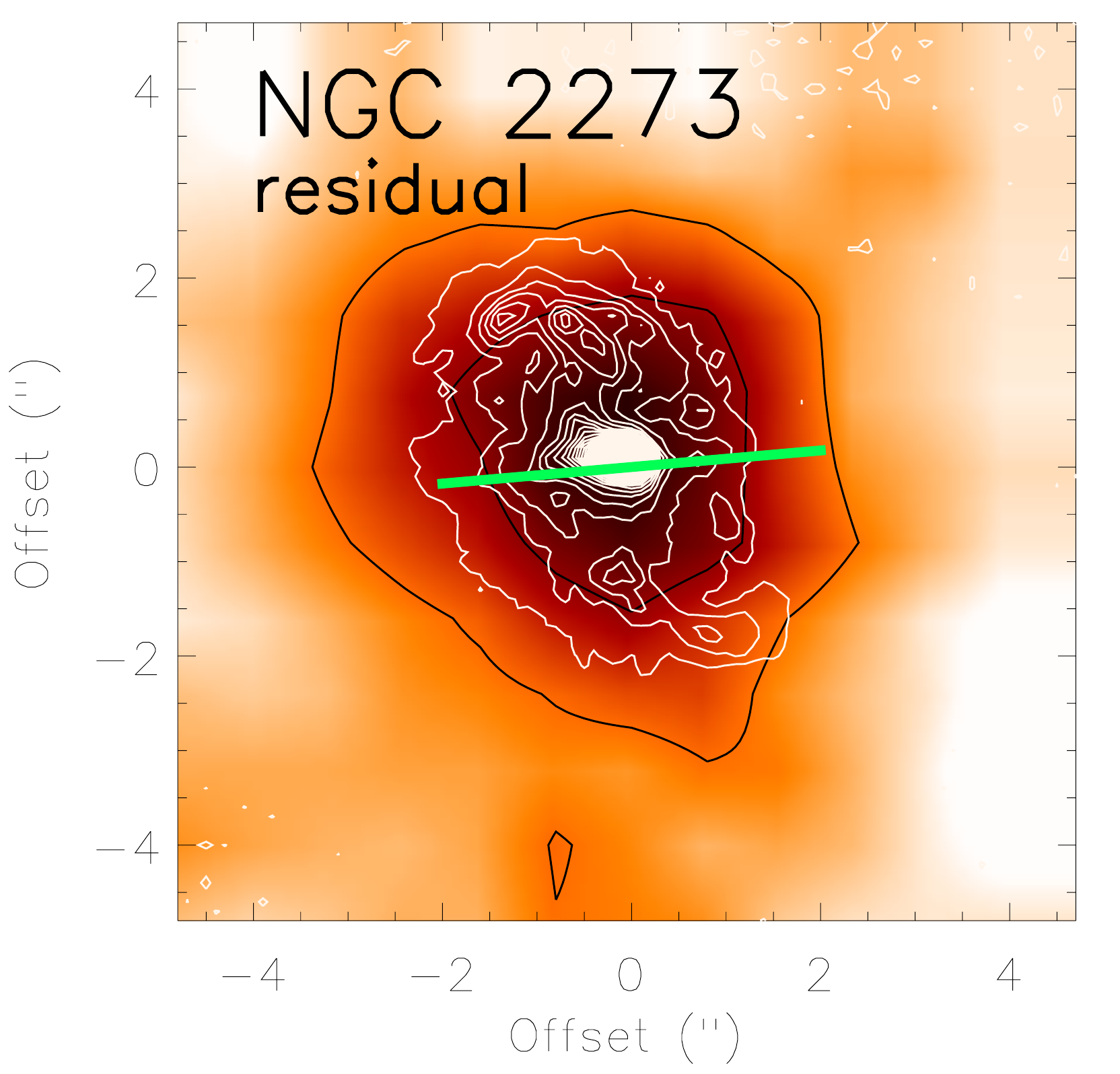}
\includegraphics[scale=0.76]{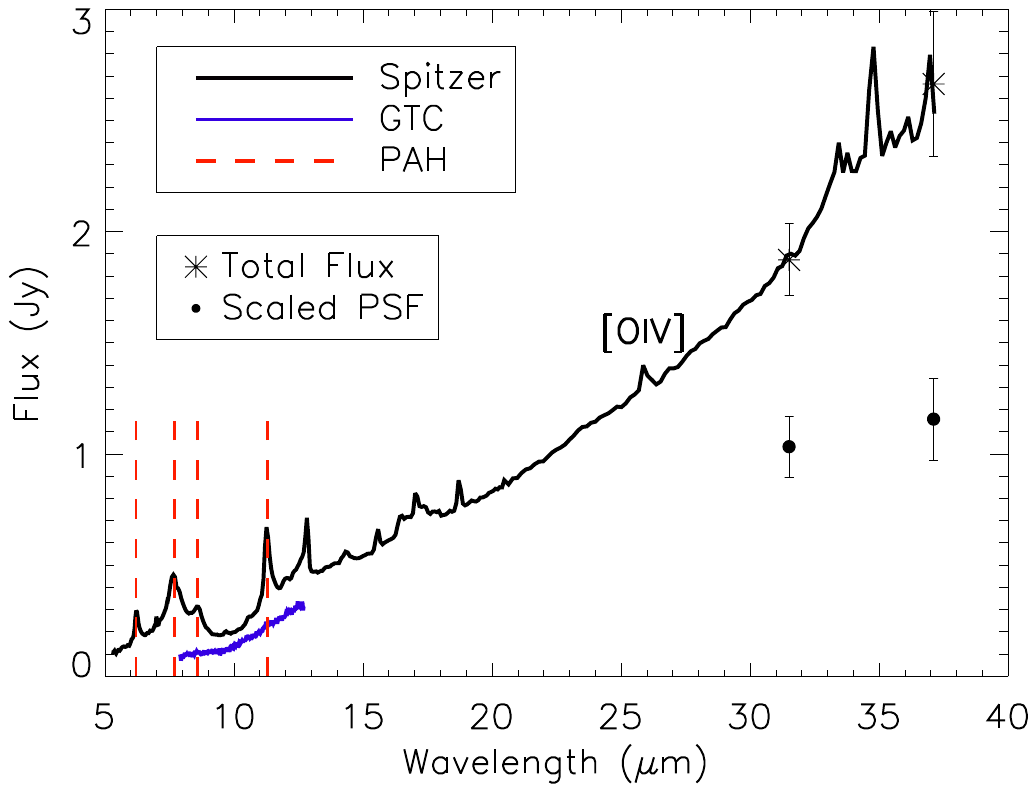}

\caption{\textit{Left}: 37.1 $\mu$m residual images of Mrk 3, NGC 1275, and NGC 2273 with 3$\sigma$ contour (black) and $HST$ optical contours overlaid in white.  The peak MIR image is centered and aligned with the peak of the optical image. The radio axis is highlighted in green at a scale of 500 pc.  In all images, North is up and East is to the left.  \textit{Right}:  $Spitzer$ spectra (black solid line) compared to sub-arcsecond N-band spectra (solid blue line).  The $Spitzer$ spectrum is also compared to the total image flux from obtained from our data (star), while the PSF-scaled flux is also shown (solid black dot).  PAH features, if present, are highlighted in red.}
\label{im_mrk3}
\end{figure*}

\subsection{Mrk 3}

The top left panel of Figure \ref{im_mrk3} shows the SOFIA 37.1 $\mu$m residual image of Mrk 3 with contours in black overlaid by optical [O \textsc{iii}] $\lambda$5007 \citep{Capetti95} contours in white.  Both wavelengths show clear elongation toward the east/west direction, suggesting a common origin.  \citet{Capetti1996} associated this optical emission with the NLR and showed a close association between NLR emission morphology and radio emission.  A large-scale radio jet system extending 2" was also seen using MERLIN \citep{Kukula93} with a P.A. along 84$^{\circ}$.  The radio axis is shown in green at a scale of 500 pc, in close alignment with the NLR ($\sim$80$^{\circ}$).  Neither the radio nor NLR axes are aligned with that of the host galaxy, at a P.A. of 15$^{\circ}$ \citep{Asmus2016}.

The top right panel of Figure \ref{im_mrk3} shows the 8 - 13 $\mu$m subarcsecond spectrum from the GTC (blue line) and the 5 - 38 $\mu$m spectrum from $Spitzer$ (black line), whose continuum emission is consistent with the total flux from SOFIA (black star).  Nuclear subarcsecond-scale emission at 12 $\mu$m accounts for $\sim$ 57\% of the arcsecond-scale emission as determined by $Spitzer$, suggesting that the extended emission lies spatially between $\sim$ 100 - 1000 pc.  The $Spitzer$ spectrum does not exhibit any notable PAH features.  However, the strong [O \textsc{iv}] 25.9 $\mu$m emission line, as well as the less prominent [Ne \textsc{v}] 24.3 $\mu$m line, indicate emission from within the NLR.  Because the spectral features from $Spitzer$ are consistent with NLR emission, and the extension in the residual image is coincident with optical and radio axes, we can conclude that the extended emission in the SOFIA residual is most likely associated with the NLR.

\subsection{NGC 1275}
\label{rg}

NGC 1275 is an atypical elliptical Seyfert galaxy displaying a network of H$\alpha$ filaments and is possibly the result of a merger \citep{Holtzman1992}.  The 37.1 $\mu$m residual image, shown on the middle left panel of Figure \ref{im_mrk3}, shows some tentative extension, possibly indicating a contribution from outflows in the galaxy \citep{Conselice2001}.  The white contours, corresponding to optical continuum emission \citep{Carlson1998}, clearly show strong, point-like emission from the nucleus, but does not show any similar extension.  The radio axis is shown in green at a PA of 160$^{\circ}$ \citep{Asmus2016}.

The $Spitzer$ spectrum in the middle right panel of Figure \ref{im_mrk3} shows almost no MIR features.  The 12 $\mu$m emission determined by the N-band GTC spectrum comprises 83\% of the total dust emission as seen by $Spitzer$, suggesting that the 10 $\mu$m silicate emission feature seen in the $Spitzer$ spectrum originates from the same subarcsecond-scale source.  For NGC 1275, this corresponds spatially to a source $\lesssim$ 140 pc.  Due to the lack of spectral lines and insufficient correspondence between wavelength axes, the data do not allow us to conclude the origin of residual emission.

\subsection{NGC 2273}

NGC 2273 is known to harbor a star forming ring within $\sim$ 2" from the AGN \citep{FWM2000,Martini2003,Sani2012}.  In the bottom left panel of Figure \ref{im_mrk3}, nuclear optical [O \textsc{iii}] $\lambda$5007 emission \citep{FWM2000} is compared with the 37.1 $\mu$m residuals from SOFIA.  GTC observations by \citet{AH2014} at 8.7 $\mu$m reveal extension to the northeast and southwest, coincident with the [O \textsc{iii}] extension in Figure \ref{im_mrk3}.  The SOFIA residual is consistent with the elongation of the [O \textsc{iii}] extensions to the southwest.  The extension is not aligned with the radio axis, PA $\sim$ 95$^{\circ}$ \citep{Nagar1999}, highlighted in green. 

The $Spitzer$ spectrum in the bottom right panel of Figure \ref{im_mrk3} shows strong PAH emission features and weak [O \textsc{iv}] 25.9 $\mu$m emission.  We conclude that the residual emission is most likely due to star formation in NGC 2273 because 1) the residual emission is approximately aligned with star forming regions, 2) there is no correlation between the radio axis and residual image, and 3) strong PAH features are seen in the $Spitzer$ spectrum.  At 12 $\mu$m, subarcsecond emission accounts for $\sim$ 63\% of the emission measured by $Spitzer$, suggesting that star formation occurs on scales $\sim$ 0.3 - 3" ($\sim$ 35 - 450 pc), within the FWHM of SOFIA.

\subsection{NGC 3081}
\label{ngc3081}

\begin{figure*}
\includegraphics[scale=0.38]{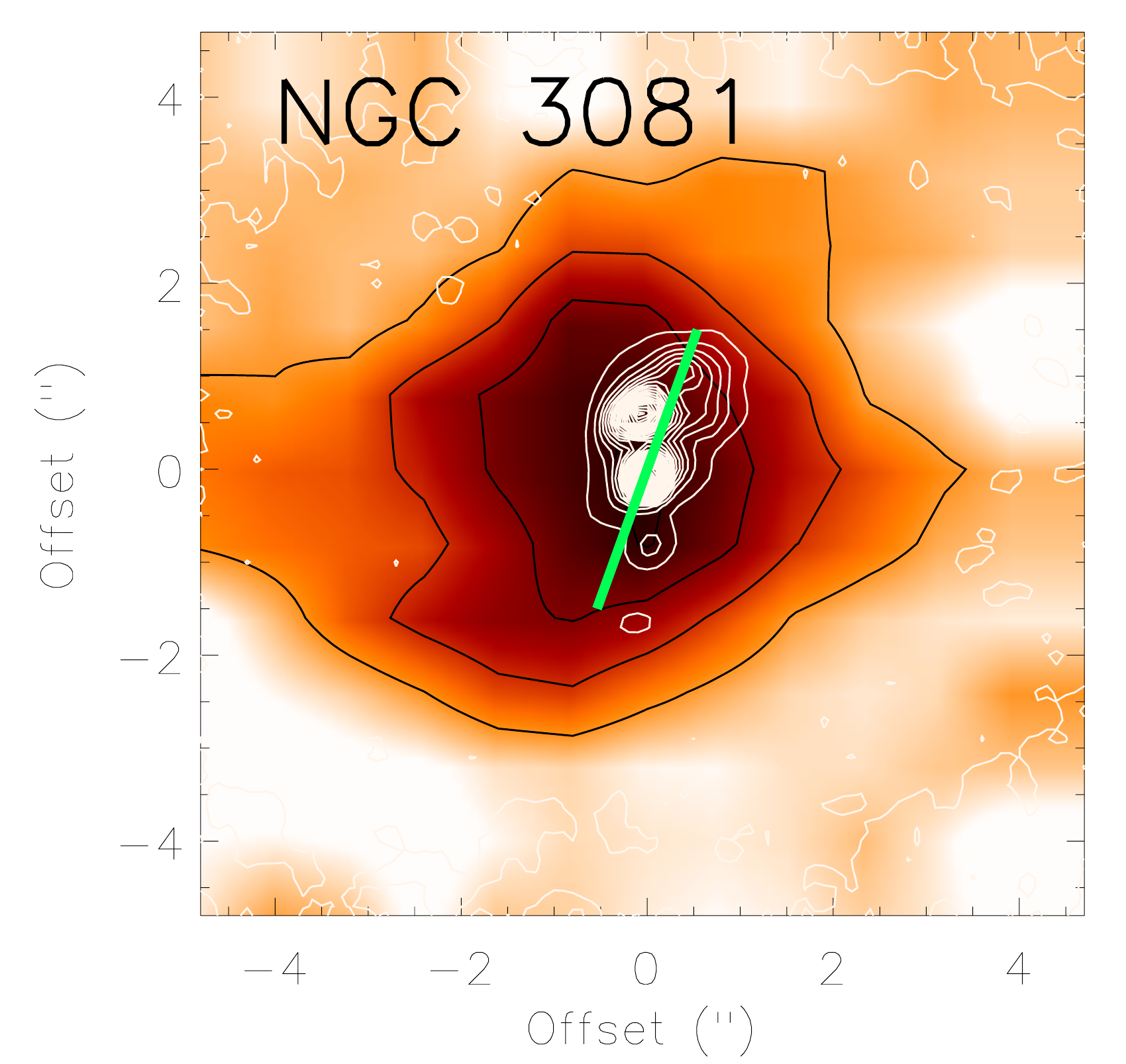}
\includegraphics[scale=0.75]{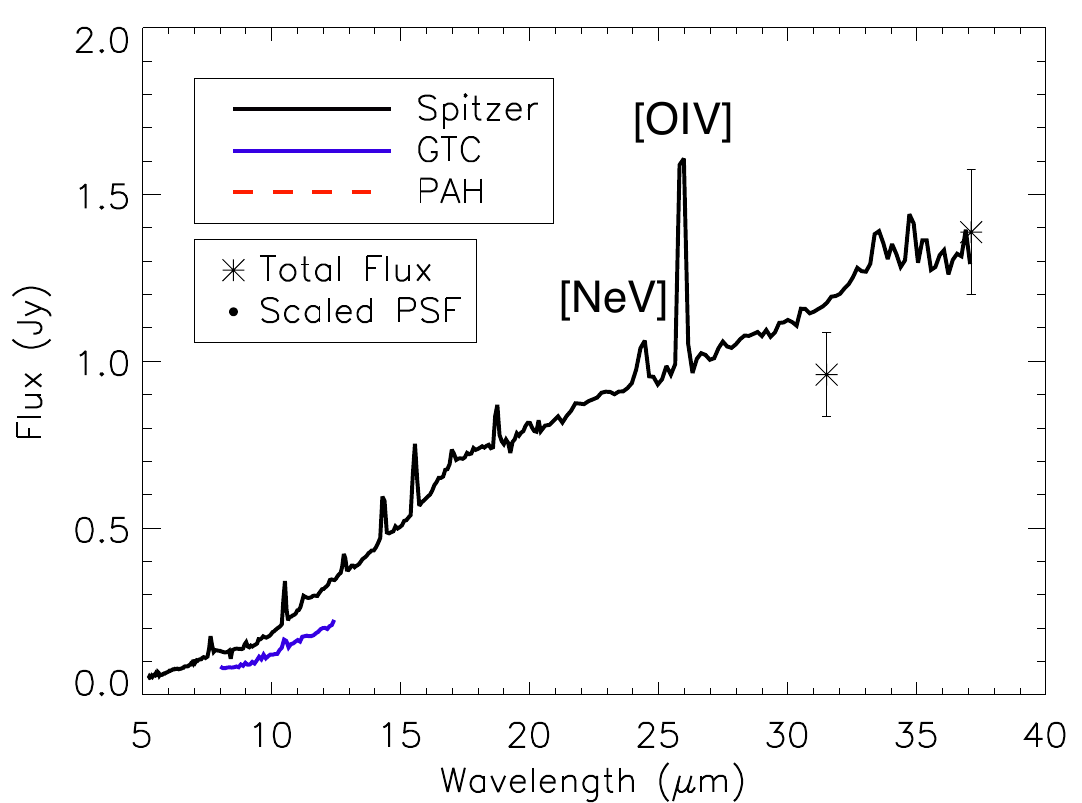}

\caption{\textit{Left}: 37.1 $\mu$m image from SOFIA with 3$\sigma$ contours (black) overlaid with optical contours (white). The peak of the MIR image is centered and aligned with the peak of the optical image.  The radio axis is highlighted in green at a scale of 500 pc.  In this image, North is up and East is to the left.  \textit{Right}: $Spitzer$ spectrum (black solid line) compared to sub-arcsecond 8 - 13 $\mu$m spectrum (solid blue line).  The $Spitzer$ spectrum is also compared to the total image flux obtained from our data (star).}
\label{im_3081}
\end{figure*}

The left panel of Figure \ref{im_3081} shows the SOFIA image of NGC 3081; the image is not included in Figure \ref{im_mrk3} because we did not perform a PSF subtraction and, hence, do not have a residual image.  On larger scales, this galaxy is notable for its series of kpc-scale diameter ringed structures: a nuclear ring ($D$ = 2.3 kpc), an inner ring ($D$ =11.0 kpc), an outer ring ($D$ = 26.9 kpc), and a "pseudo-ring" ($D$ = 33.1 kpc) \citep{Buta1990,Buta1998,Buta2004,Byrd2006}.  The nuclear ring has a PA of $\sim$120$^{\circ}$ and a major axis of $\sim$ 12" \citep{FWM2000}. On smaller scales, [O \textsc{iii}] $\lambda$5007 observations clearly show a bright region of optical emission $\sim$ 1" north of the nucleus \citep{FWM2000}.  The [O \textsc{iii}]/([N \textsc{ii}]+H$\alpha$) ratio of the bright emission is similar to that of the nucleus suggesting that the bright region is related to dust or gas heated by the AGN and that stellar processes are not responsible for photoionization in that region.  \citet{RA2011} presented $Herschel$ imaging data from 70 to 500 $\mu$m, where they fitted the NIR to FIR SED and concluded that on scales $\leq$ 0.85 kpc, the FIR nuclear luminosity was reproduced by cool dust in the torus heated by the AGN.

\citet{Nagar1999} observed a radio axis $\sim$160$^{\circ}$ without a prominent radio jet.  Neither the optical nor radio axes are consistent with the east-west extension seen in the SOFIA 37.1 $\mu$m image, giving an unclear explanation to its origin.  This inconsistency further compelled the use of total flux from the PSF scaling as an upper limit, as mentioned in Section \ref{imaging}.

The $Spitzer$ spectrum in Figure \ref{im_3081} shows very weak PAH emission, but does show very strong [O \textsc{iv}] 25.9 $\mu$m emission line, as well as [Ne \textsc{v}] 24.3 $\mu$m emission line, suggesting NLR activity.  At 12 $\mu$m, subarcsecond emission as determined by Gemini/T-ReCS accounts for $\sim$65\% of emission from $Spitzer$, signaling that any extended emission occurs on scales $\sim$ 50 - 600 pc.  The $Spitzer$ continuum is consistent with the total SOFIA emission at 37.1 $\mu$m, however it is not consistent with the 31.5 $\mu$m total flux.

\subsection{NGC 3227}

NGC 3227 is another galaxy known for circumnuclear star formation \citep{Schinnerer2001,RA2003,Davies2006}.  The top left panel of Figure \ref{im_3227} shows the 37.1 $\mu$m residual emission, as well as optical emission \citep{Malkan1998} using the F606W $HST$ filter.  \citet{Mundell1995} found that the optical [O \textsc{iii}] $\lambda$5007 axis has a P.A. $\sim$ 30$^{\circ}$ extending $\sim$ 500 pc ($\sim$ 6.5") to the NE that is aligned with the NLR, but does not coincide with the radio axis (PA $\sim$ -10$^{\circ}$).  We do not find a similar extension in the SOFIA observations.  However, \citet{Schmitt1996} found using $HST$ observations that the [O \textsc{iii}] $\lambda$5007 emission extends to the NE with P.A. $\sim$ 15$^{\circ}$ at a distance of 100 pc (1.4"), thus within the FWHM of our SOFIA observations.

The $Spitzer$ spectrum on the top right panel of Figure \ref{im_3227} shows prominent PAH features. Comparing that to the subarcsecond GTC spectrum, which shows no PAH features, suggests that most star forming activity occurs between 0.3 - 3 arcseconds ($\sim$ 20 - 270 pc), similar to the comparison of VISIR to $Spitzer$ by \citet{Honig2010,Jensen2017}. \citet{Schinnerer2001} found a nuclear stellar cluster within $\sim$ 70 pc of the core.  The 12 $\mu$m emission estimated from the GTC spectrum accounts for $\sim$60\% of arcsecond-scale emission as determined by $Spitzer$.  The spectrum also shows that the [O \textsc{iv}] 25.9 $\mu$m line, while not as prominent as in other AGN in this sample, is still clearly detected.  While the [O \textsc{iv}] feature would suggest activity in the NLR, strong PAH in the spectrum suggests that the source of residual emission is primarily stellar heating.  Both emission sources should be considered when modeling the mid- to far-IR emission of this AGN.

\subsection{NGC 4151}

NGC 4151 is a nearby, well-studied Seyfert 1.5 galaxy.  The residual image is shown in the middle left panel of Figure \ref{im_3227} with optical contours \citep{Kaiser2000} overlaid in white.  Using MIR observations from Gemini, \citet{Radomski2003} showed extended emission at 10.8 and 18.2 $\mu$m in the central 3.5" at a P.A. of $\sim$ 60$^{\circ}$.  They demonstrated that the MIR extension coincides with the NLR, as determined by [O \textsc{iii}] $\lambda$5007 observations \citep{Evans1993,Kaiser2000}.  Additionally, \citet{Radomski2003} very thoroughly isolated the extended emission as NLR dust based on grain size and composition.  While the comparison between SOFIA residuals and HST contours in Figure \ref{im_3227} are not quite on the same scale, the general P.A. $\sim$ 67$^{\circ}$ of the [O \textsc{iii}] $\lambda$5007 emission \citep{Asmus2016} is roughly consistent with the extension in the SOFIA residual, and is similar to the radio axis P.A. $\sim$ 77$^{\circ}$ \citep{Pedlar1998}, shown in green at a scale of 500 pc.

The $Spitzer$ spectrum in the middle right panel of Figure \ref{im_3227} shows [O \textsc{iv}] and [Ne \textsc{v}] emission lines, with minimal PAH, indicative of little or no major star formation.  Comparison of subarcsecond-scale emission at 12 $\mu$m shows that nuclear emission accounts for $\sim$ 70\% of emission determined by $Spitzer$, suggesting that extended emission occurs between $\sim$ 20 - 225 pc.  The consistency of the P.A. of the 37.1 $\mu$m image with optical line emission, as well as arcsecond-scale spectral features, strongly suggests that the source of extended emission in NGC 4151 is the NLR.

\subsection{NGC 4388}

\begin{figure*}

\includegraphics[scale=0.4]{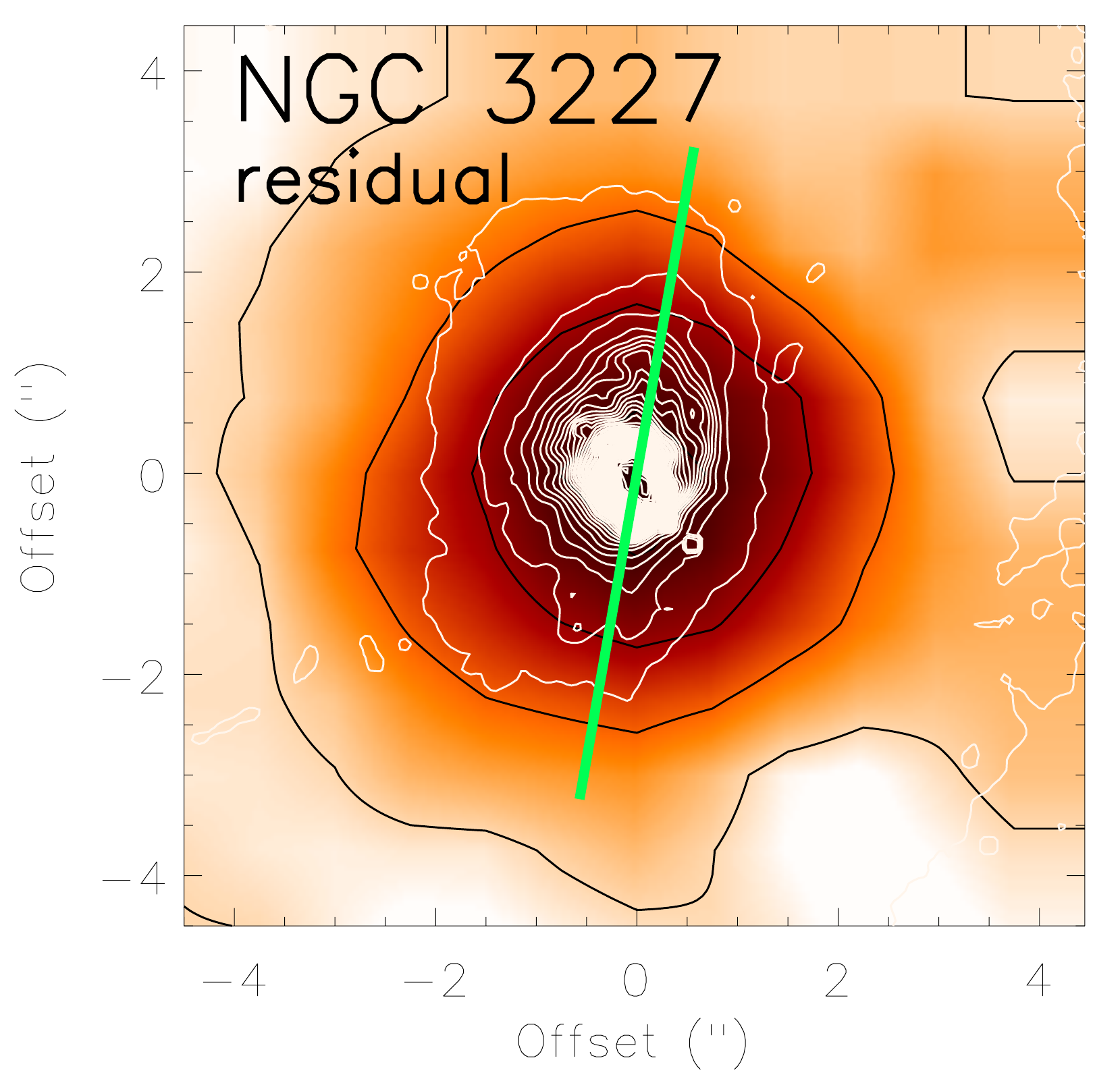}
\includegraphics[scale=0.8]{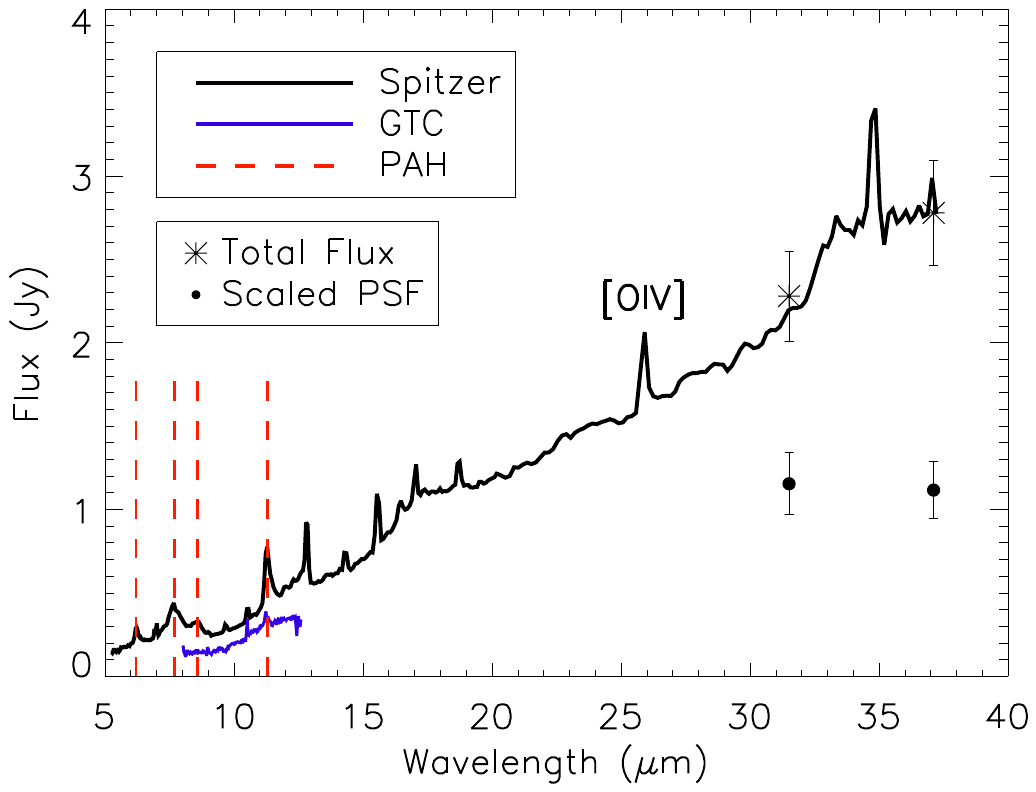}

\includegraphics[scale=0.4]{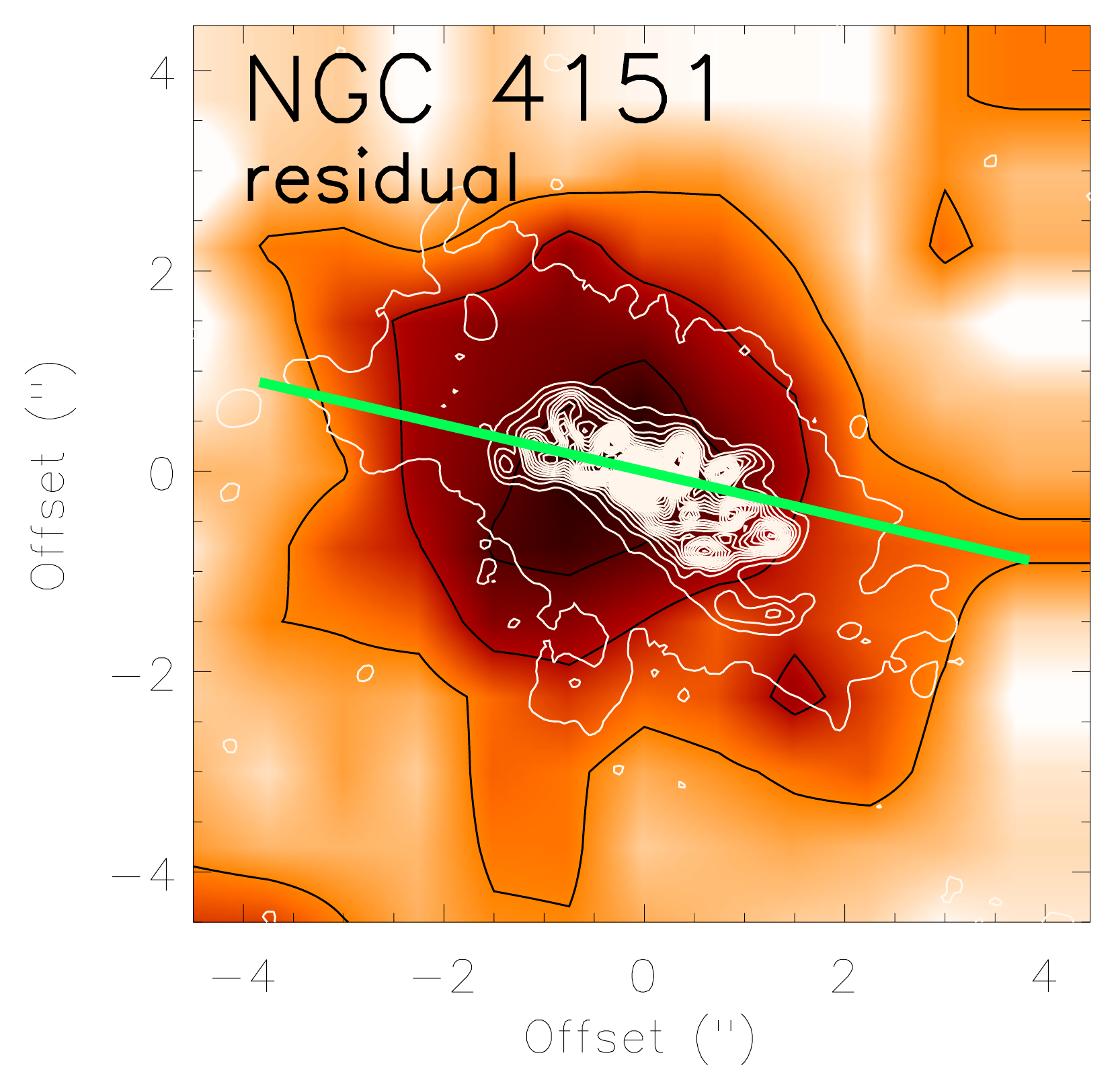}
\includegraphics[scale=0.8]{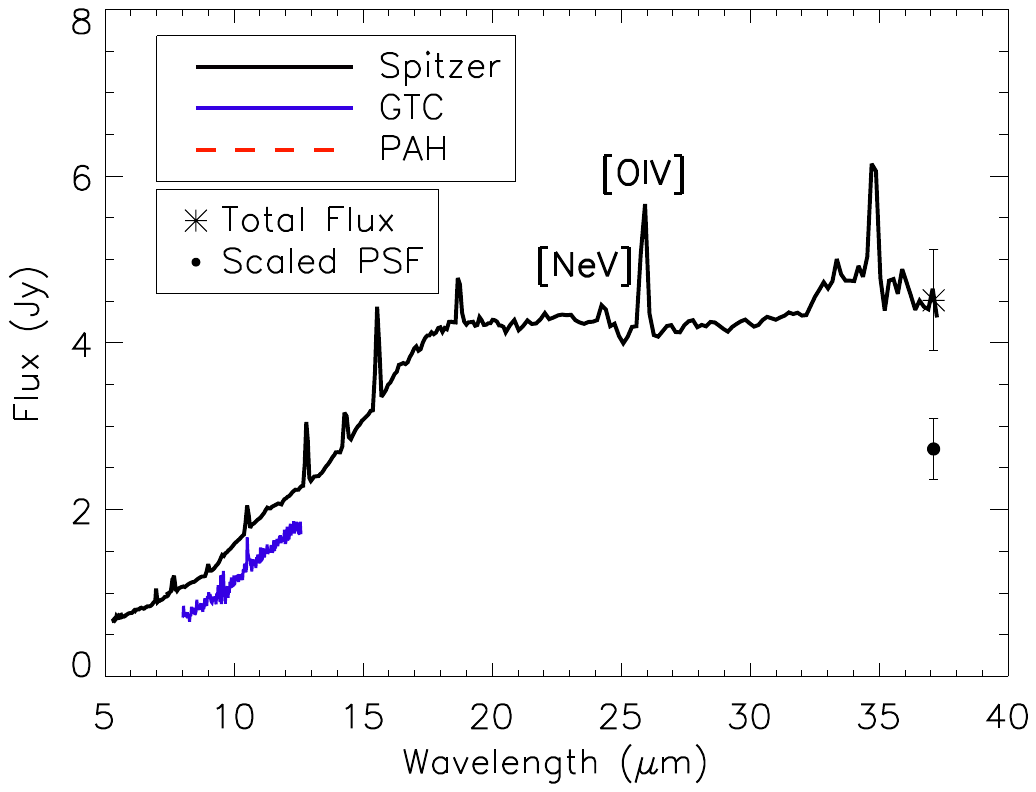}

\includegraphics[scale=0.39]{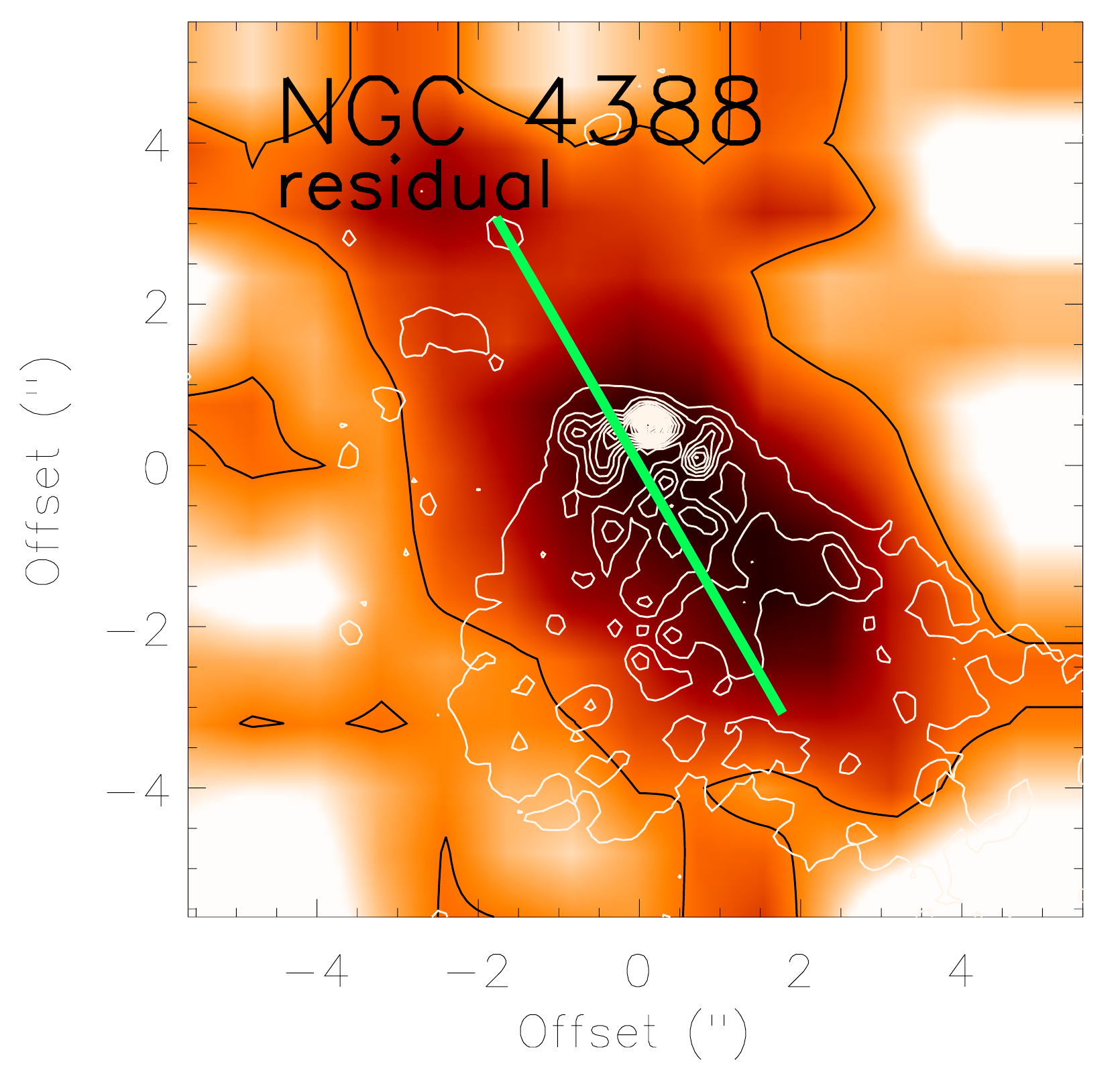}
\includegraphics[scale=0.79]{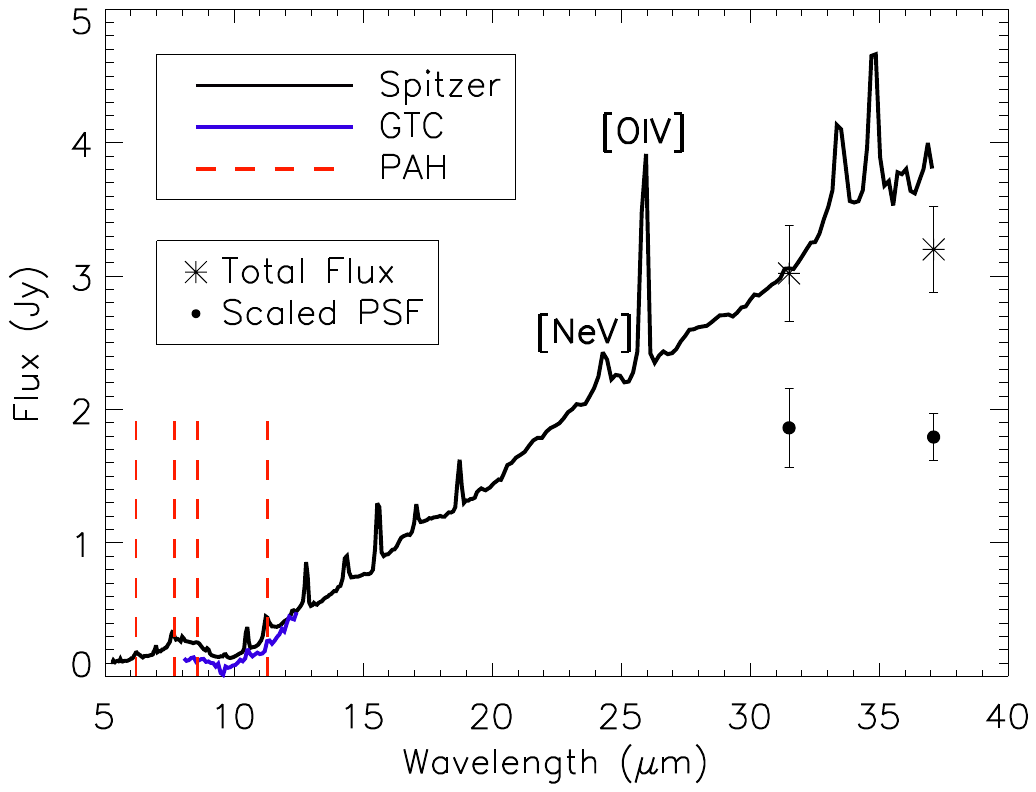}

\caption{\textit{Left}: 37.1 $\mu$m residual images with 3$\sigma$ contours (black) with optical contours overlaid in white.  The peak of the MIR image is centered and aligned with the peak of the optical image. The radio axis is highlighted in green at a scale of 500 pc.  In all images, North is up and East is to the left.  \textit{Right}: $Spitzer$ spectra (black solid line) compared to sub-arcsecond 8 - 13 $\mu$m spectra (solid blue line).  The red dashed vertical lines highlight the strength of PAH features.  The $Spitzer$ spectra are also compared to the total image flux obtained from our data (star), while the PSF-scaled flux is also shown (solid black dot).}
\label{im_3227}
\end{figure*}

The 37.1 $\mu$m residual image of NGC 4388 in the bottom left panel of Figure \ref{im_3227} clearly shows extension to the northeast and southwest at P.A. $\sim$ 40$^{\circ}$, in the same direction as well-known galactic outflows \citep{Veilleux1999,RA2017}.  The optical [O \textsc{iii}] $\lambda$5007 white contours \citep{Falcke1998} show a well-defined ionization cone to the southwest.  The northeast extension, clearly visible in the residual image, is likely extinguished in the optical image by the host galaxy, which has a P.A. of 91$^{\circ}$.  The good alignment between the optical and radio morphologies (P.A. 30$^{\circ}$) suggests that the radio jet in NGC 4388 interacts with gas in the NLR \citep{Falcke1998}.  NIR spectroscopic observations of \citet{RA2017} show that photoionization by radiation from the central engine does not account for all photoionization observed in the ionization cone and suggest that dust interaction with the radio jet must occur in the central few hundred parsecs of the nucleus.    

The $Spitzer$ spectrum in the bottom right panel of Figure \ref{im_3227} shows some PAH emission, though it is possible that the 7.7 $\mu$m feature is blended with [Ne \textsc{vi}] 7.6 $\mu$m.  The [O \textsc{iv}] 25.9 $\mu$m line is prominent and is neighbored by a visible [Ne \textsc{v}] fine structure line. Subarcsecond-scale 12 $\mu$m emission as determined by the GTC spectrum accounts for $\sim$ 88\% of 12 $\mu$m $Spitzer$ emission, suggesting a common source found spatially $\lesssim$ 40 pc.  However, the PSF scaling suggests that this percentage decreases at longer wavelengths.  Even though this spectrum shows some star forming activity, the distinct alignment of the residual image to the optical and radio axes suggests that the extended 37.1 $\mu$m emission is primarily from the NLR, while star formation may have some minor contribution. 

\subsection{Extended NLR Emission}
\label{dusty}

Figures \ref{im_mrk3} and \ref{im_3227} of Mrk 3 and NGC 4388 clearly show extended emission in the direction of the NLR and radio axes, as well as spectra that are consistent with NLR emission.  The spectrum of NGC 4151 is also consistent with NLR emission, but the residual image axis is not quite as clear.  This could be because Mrk 3 and NGC 4388 are both Sy2 AGN and seen edge-on, whereas NGC 4151 is a Sy1.5.  According to \citet{Fischer2013}, the inclination angle of the NLR bicone of Mrk 3 is 5$^{\circ}$ whereas the inclination of the NLR bicone of NGC 4151 is 45$^{\circ}$, indicating that clear elongation would not be seen in the residual image of NGC 4151.  

Table \ref{residual} shows that the residual fluxes for the given AGNs in our sample increase from 31.5 to 37.1 $\mu$m (NGC 3081 and NGC 4151 are not included due to lack of data).  From the residual fluxes, a tentative estimate of the temperature for the emitting regions can be made by modeling the emission as a single temperature blackbody.  Temperatures given in the table are those that show a least squares fit to the residual fluxes, and the errors indicate the range of temperatures which fit within their error bars.  The uncertainty in the measurement would be greatly improved with more data points. 

Using these tentative dust temperatures, the dust mass can be estimated by the relation:
\begin{equation}
F_{\nu} = \kappa_{\nu} B_{\nu}(T) M_{d} D^{-2}_{L}
\end{equation}

\citep{Casey2012}, where $F_{\nu}$ used is the 37.1 $\mu$m residual flux in Table \ref{residual}, $\kappa_{\nu}$ is the dust opacity \citep{Li2001}, $B_{\nu}(T)$ is the blackbody function using $\lambda$ = 37.1 $\mu$m and the temperatures in Table \ref{residual}, and $D$ is the distance given in Table \ref{info}.  The estimated dust masses given in Table \ref{residual} range between $\sim$ 10$^{3}$ - 10$^{5}$ M$_{\sun}$.  This is only representative of dust directly heated by the AGN, and does not include any dust heated by star formation.  The dust mass in NGC 1275 is at least an order of magnitude greater than the dust masses in the remainder of the sample.  This is the only radio galaxy in our sample, and it has a well-known galactic outflow, as mentioned in Section \ref{rg}.

Using the same relation, \citet{GB2014} found the dust mass in the NLR of NGC 1068 to be $M_{\rm dust}$ = (8 $\pm$ 2) $\times$ 10$^{5}$ M$_{\sun}$ in a 400 $\times$ 200 pc region.  For a large sample of Seyferts, \citet{Ho2009} found ionized gas mass in a 400 $\times$ 200 pc region of the NLR to be $M_{\rm NLR} \sim$ 3 $\times$ 10$^{4}$ M$\sun$, though \citet{Vaona2012} find an ionized gas mass of $\sim$ 10$^{6-8}$ M$_{\sun}$ in larger radii.  Assuming the Galactic gas to dust ratio ($\sim$ 10$^{2}$), NLR dust masses could range from 10$^{2-6}$ M$_{\sun}$.

\begin{table}
\begin{minipage}{65mm}
\centering
\caption{31.5 and 37.1 $\mu$m residual fluxes}
\tiny{
\begin{tabular}{ccccc}

\hline
\hline
Object 	&	31.5 $\mu$m Res.	&	37.1 $\mu$m Res.	&	Estimated 	&	Dust Mass	\\
		&	flux (Jy)				&	flux (Jy)				&	Temp. (K) &	(M$_{\sun}$)	\\
\hline		
Mrk 3 	&	1.1$\pm$.4			&	1.2$\pm$.5			&	78$^{+90}_{-19}$			&	1.1$^{+7.4}_{-0.4}$$\times$10$^{4}$		\\
NGC 1275&	1.0$\pm$.7			&	1.8$\pm$.7			&	51$^{+23}_{-8}$			&	4.0$^{+13}_{-1.5}$$\times$10$^{5}$		\\
NGC 2273&	0.8$\pm$.2			&	1.5$\pm$.4			&	51$^{+12}_{-6}$			&	4.9$^{+15}_{-1.5}$$\times$10$^{4}$		\\
NGC 3081&	\dots					&	\dots					&	\dots						&	\dots			\\
NGC 3227&	1.1$\pm$.3			&	1.7$\pm$.4			&	59$^{+13}_{-7}$			&	6.4$^{+15}_{-2.3}$$\times$10$^{3}$		\\
NGC 4151&	\dots	 				&	1.8$\pm$.8			&	\dots						&	\dots			\\
NGC 4388&	1.2$\pm$.5			&	1.4$\pm$.4			&	70$^{+32}_{-13}$			&	2.8$^{+10}_{-1.0}$$\times$10$^{3}$		 \\	

\hline
\hline
\end{tabular}}

\label{residual}
\end{minipage}
\end{table}

Considering dust grains such as silicates that are directly heated by the central engine with no intervening dust, \citet{Gratadour2006} simplified the calculation of \citet{Barvainis1987} to describe the temperature, $T$, of dust grains at a distance, $r$, from the nucleus given a UV luminosity, $L_{uv,46}$,

\begin{equation}
T = 1650 \times \left(\frac{L_{\rm uv,46}}{r^{2}_{\rm pc}}\right)^{\frac{1}{5.6}} K
\end{equation} 

in order to put an upper limit on the temperature caused by AGN heating.  Here, we use the blackbody temperatures in Table \ref{residual} to estimate the distance at which dust can be heated by the central excitation source. \citet{Kishimoto2002} find an upper limit for the UV luminosity of Mrk 3 as $\sim$ 3 $\times$ 10$^{44}$ erg s$^{-1}$, resulting in a radial distance of $r$ $\sim$ 900 pc.  The extent of the NLR for Mrk 3 according to Figure \ref{im_mrk3} is about 4", which corresponds to $\sim$ 1 kpc in the SOFIA image.  \citet{Colina1992} report a UV luminosity of 1.7 $\times$ 10$^{43}$ erg s$^{-1}$ for NGC 4388, suggesting that dust can be heated to a distance $\sim$ 300 pc.  The extended emission shown in Figure \ref{im_3227} reaches $\sim$ 350 pc.  While the relation underestimates the radial extent that the central excitation source given the estimated temperatures, we show here that the extended emission seen in the residual images is consistent with spatial scales at which dust can be heated by the central engine (0.1 - 1 kpc).  These physical scales are consistent with other studies \citep{Groves2006,Schweitzer2008,Mor2009,Mor2012}.  More data regarding the dust temperature is needed to refine this analysis.

\subsection{MIR $Spitzer$ Continuum}

\citet{Deo2009} showed that Seyfert 2 AGN can be separated into two groups - those that show strong PAH in their $Spitzer$ spectra, and those that are AGN-dominated and show little to no PAH.  Those that are AGN-dominated have much flatter continua between 20 - 30 $\mu$m.  Likewise, we find that Mrk 3 and NGC 4151 have very weak traces of PAH in their spectra, strong [O \textsc{iv}] emission, and also have the flattest 20 - 30 $\mu$m continua in our sample.  These two AGN also show collimation along the system axis in the 37.1 $\mu$m residual image.  This suggests that AGN with flat 20 - 30 $\mu$m $Spitzer$ continua would likely contain an NLR component in their MIR SED.


\section{Nuclear SED Modeling}
\label{bayes}

In order to analyze the nuclear IR SED of the unresolved torus, we compiled the highest angular resolution NIR and MIR data available from the literature, for the 7 AGN in our sample. Subarcsecond-resolution NIR fluxes are available for 6 galaxies using NICMOS/\textit{HST} and IRCam3/UKIRT.  Mrk 3 is an exception, wherein we use NIR flux data from \textit{WISE} \citep{Ichikawa2017} as an upper limit.  Subarcsecond-resolution MIR photometry was obtained for 6 galaxies using Gemini North/South, Subaru, and VLT; MIR \textit{N} and \textit{Q} band photometry for NGC 2273 was not available.   N-band observations probe the central $\sim$ 0.3" - 0.6", which corresponds to $\sim$ 20 - 35 pc for NGC 2273, NGC 3227, NGC 4151, and NGC 4388.  However, for Mrk 3, NGC 1275, and NGC 3081, this resolution probes scales $\sim$ 90 - 140 pc.  For all objects at all wavelengths, non-torus contaminating emission was estimated and removed in their respective analyses by the authors of the papers cited in Table \ref{photometry}.

In the previous section we showed that some extended emission from SOFIA observations is due to a component that we suggest is dust in the NLR.  An accurate estimation of the level of contribution to the total SED is highly model dependent, but we note that when modeling mid- to far-IR emission, both star formation and the additional component should be taken into consideration.  For this reason, and because of the inability to perform a spectral decomposition as explained in Section \ref{imaging}, we use the PSF-subtracted photometric data from SOFIA as an upper limit to model the 1 - 40 $\mu$m SED of the unresolved torus.

The \textsc{Clumpy} torus models of \citet{Nenkova2008a} assert that dust around the central engine of an AGN is distributed in clumps that can be primarily described by six physical properties: a radial distribution power law index $q$ ($\propto$ $r^{-q}$); the torus opening angle width $\sigma$; the torus inclination angle $i$;  total number of clouds in some line of sight $N_{0}$; an optical depth per cloud, $\tau$$_{v}$; and the outer to inner radial ratio $Y=R_{\mbox{\tiny out}}$/$R_{\mbox{\tiny in}}$ where $R_{\mbox{\tiny in}}$ is set set by the dust sublimation temperature, $T_{\mbox{\tiny sub}}$$\sim$1500K, and computed by bolometric luminosity $R_{\mbox{\tiny in}}$ = 0.4($L_{\mbox{\tiny bol}}$/$10^{45}$)$^{0.5}$ pc \citep{Barvainis1987}.

Using the Bayesian inference tool \textsc{BayesClumpy} \citep{AR2009}, we fit the IR SED using photometry in Table \ref{photometry} and spectroscopy in Table \ref{spectroscopy} as inputs to infer physical properties of the torus.  This tool uses the photometric and spectroscopic inputs to detect all possible combinations of model parameters consistent with the observations, and outputs probability distributions of the parameters.  We used the highest resolution photometric and spectroscopic data available to construct the 1-20 $\mu$m SEDs, then added SOFIA data from \citetalias{F16} and this work.  The spectra have been resampled to $\sim$ 50 data points, following the methodology of \citet{AH2013,RA2014}.

\begin{table*}
\centering
\begin{minipage}{160mm}

\caption{High spatial resolution NIR and MIR flux data}
\footnotesize{
\begin{tabular}{ccccccccc}

\hline
Object	&	&	&	&	Flux Densities (mJy)	 \\
	&	\textit{J} &	\textit{H} 	&	\textit{K}	&	\textit{L}	&	\textit{M}	&	\textit{N}	&	\textit{Q}	&	Ref(s).	\\
\hline
Mrk 3		&	\dots		&	\dots		& \dots	& <0.06	& <0.076 &	448 $\pm$ 120 &	\dots	& a,b 	\\
NGC 1275	&	 \dots	&  4.3 $\pm$ 0.4	& \dots &	\dots	&	\dots		&	 886 $\pm$ 11	&	\dots		&	c,b	\\

NGC 2273	&	\dots		& 0.32 $\pm$ 0.28	& \dots &	\dots	&	\dots		&	\dots	&	\dots		& c	\\

NGC 3081	&	\dots		& 0.22 $\pm$ 0.13	& \dots&	 \dots	& \dots		&	83 $\pm$ 12	&	231 $\pm$ 58	& c,d 	\\

NGC 3227	&	\dots	& 7.8 $\pm$ 0.8&	16.6 $\pm$ 1.7	&	46.7 $\pm$ 9.3	&	72 $\pm$ 27	& 180 $\pm$ 11	& 772 $\pm$ 47 &	  e,f,g,h,b \\
			&		&			&				&				&				& 320 $\pm$ 22 & 	& h\\
			&		&			&				&				&				& 401 $\pm$ 60 &    &	d\\

NGC 4151	&	60 $\pm$ 3 & 100 $\pm$ 5 & 197 $\pm$ 10 &	325 $\pm$ 65	& 449 $\pm$ 34 &  1320 $\pm$198 & 3200 $\pm$ 800 & e,g,d \\

NGC 4388	&	0.06 $\pm$ 0.02 & 0.71 $\pm$ 0.28 & \dots 	& 40 $\pm$ 8 	&	\dots	 &	195 $\pm$ 29	&	803 $\pm$ 201 &  f,d 	\\
\hline
\end{tabular}\\}
 
\textsc{References:} a) \citet{Ichikawa2017} b) \citet{Asmus2014}, c) \citet{Quillen2001}, d) \citet{RA2009}, e) \citet{Kishimoto2007}, f) \citet{AH2003}, g) \citet{Ward1987}, h) \citet{Honig2010}.
\textsc{Notes.} The \textit{N} band photometry for NGC 3227 was taken in three different filters: 8.99, 11.88, and 11.29 $\mu$m using VLT and Gemini. 
\label{photometry}
\end{minipage}
\end{table*}

\begin{figure*}

\includegraphics[scale=0.7]{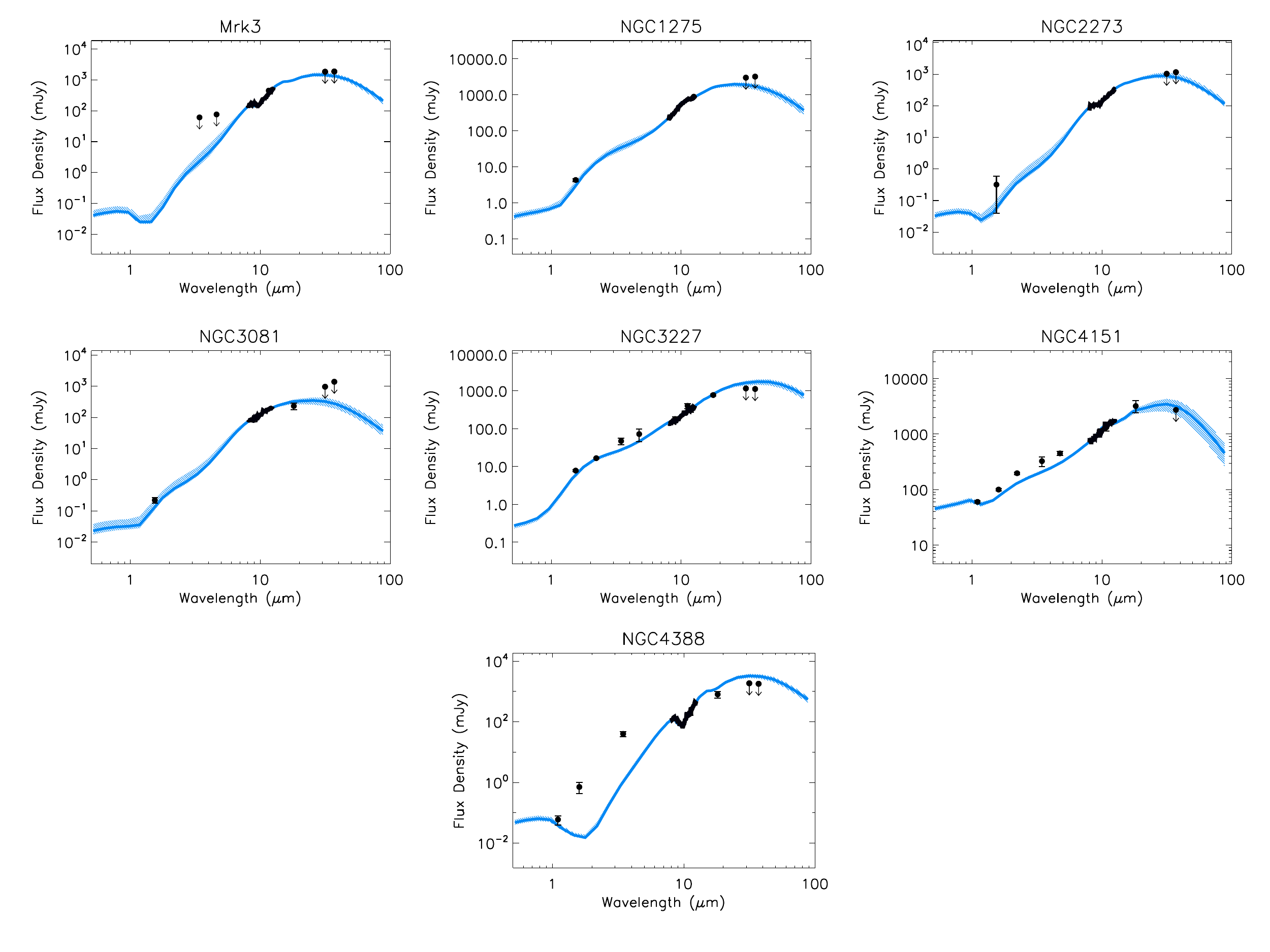}
\caption{\textsc{BayesClumpy} torus model fits.  The blue line indicates the output SED computed with the median value of the probability distribution of each parameter.  The blue shaded region indicates the range of models compatible with a 68\% confidence level around the median.  The output SEDs are shown with photometric (black dots) and 8 - 13 $\mu$m spectroscopic (black line) inputs.}
\label{SEDs}
\end{figure*}

The output model SEDs are shown in Figure \ref{SEDs}.  The blue lines mark the output SEDs computed with the median values while the blue shaded regions indicate 1$\sigma$ uncertainty.  Photometric and spectroscopic inputs are shown in black.  The SED results suggest that we may have observationally begun to determine a range of wavelengths where peak emission from the torus occurs, though the difference in flux densities is within error (see Table \ref{flux}).  Likewise, \citet{ELR2018} find a similar peak wavelength of MIR emission for NGC 1068 between 30 - 40 $\mu$m.

Even though the 31.5 and 37.1 $\mu$m photometry is given as an upper limit, the model overestimates torus emission in NGC 3227 and NGC 4388.  \citetalias{F16} show that the model SEDs tend to significantly overestimate torus emission in the absence of 31.5 $\mu$m data.  The overestimation of the SED in Figure \ref{SEDs} may suggest that a simple clumpy dust model may not fully describe torus emission, and perhaps a two-phase medium \citep[e.g.][]{Siebenmorgen2015} should be explored in more detail.  Figure \ref{SEDs} shows a poor NIR fitting to NGC 4388, and the 8 - 13 $\mu$m spectrum also shows a deep silicate absorption feature.  It is possible that a dusty NLR contaminates the NIR data and  
increases the amount of silicate absorption.

\begin{table*}

\begin{minipage}{100mm}
\centering
\caption{\textsc{BayesClumpy} output model parameters.}
\begin{tabular}{cccccccc}
\hline
Object	&	$\sigma$		&	$Y$		&	$N_{\mbox{\tiny 0}}$		& 	$q$		&	$\tau_{\mbox{\tiny V}}$	&	$i$	& $R_{\mbox{\tiny out}}$ 	\\
		&	[15$^{\circ}$,70$^{\circ}$]	&	[5,30]	&	[1,15]	&	[0,3]	&	[5,150]	&	[0$^{\circ}$,90$^{\circ}$]	& (pc)					\\
\hline
Mrk 3 	&	57$_{-5}^{+7}$	&	11$_{-1}^{+2}$	&	14$_{-1}^{+1}$		&	0.54$_{-0.35}^{+0.57}$&	39$_{-4}^{+9}$	  &	81$_{-8}^{+5}$ & 5.2 $_{-0.5}^{+0.9}$	\\
				
NGC 1275 &	26$_{-5}^{+5}$ &	10$_{-1}^{+2}$ 	&	13$_{-1}^{+1}$		& 1.17$_{-0.60}^{+0.47}$	&	144$_{-8}^{+4}$ &	75$_{-6}^{+5}$	& 3.1 $_{-0.3}^{+0.6}$	\\
		 
NGC 2273 &	62$_{-6}^{+4}$ &	15$_{-4}^{	+7}$	&	13$_{-2}^{+1}$		& 1.91$_{-0.29}^{+0.24}$	&	 78$_{-6}^{+6}$  &  78$_{-11}^{+7}$ & 1.7 $_{-0.5}^{+0.8}$	\\
		 
NGC 3081 &	62$_{-7}^{+4}$ 	&	12$_{-5}^{+10}$&	12$_{-2}^{+2}$		& 2.48$_{-0.73}^{+0.31}$	&	106$_{-19}^{+19}$ &	59$_{-24}^{+18}$ & 1.8 $_{-0.8}^{+1.5}$ 	\\
		 
NGC 3227 &	56$_{-2}^{+3}$ &	19$_{-1}^{+1}$	&	13$_{-2}^{+1}$		& 0.03$_{-0.02}^{+0.04}$	&	149$_{-2}^{+1}$.  &	10$_{-6}^{+6}$  &1.0 $_{-0.1}^{+0.1}$ 	\\
		
NGC 4151 &	15$_{-1}^{+1}$	&	26$_{-4}^{+2}$	&	10$_{-2}^{+2}$		& 1.55$_{-0.08}^{+0.12}$	&	133$_{-12}^{+10}$	&	70$_{-1}^{+1}$	& 2.2 $_{-0.3}^{+0.2}$ \\
		 
NGC 4388 &	68$_{-2}^{+1}$	&	23$_{-1}^{+2}$	&	15$_{-1}^{+1}$		& 0.60$_{-0.20}^{+0.27}$	&	31$_{-2}^{+3}$	&	  85$_{-5}^{+3}$ & 6.4 $_{-0.3}^{+0.6}$	\\

\hline
\end{tabular}
\label{parameters}
\end{minipage}

\end{table*}

\begin{figure*}

\includegraphics[scale=0.65]{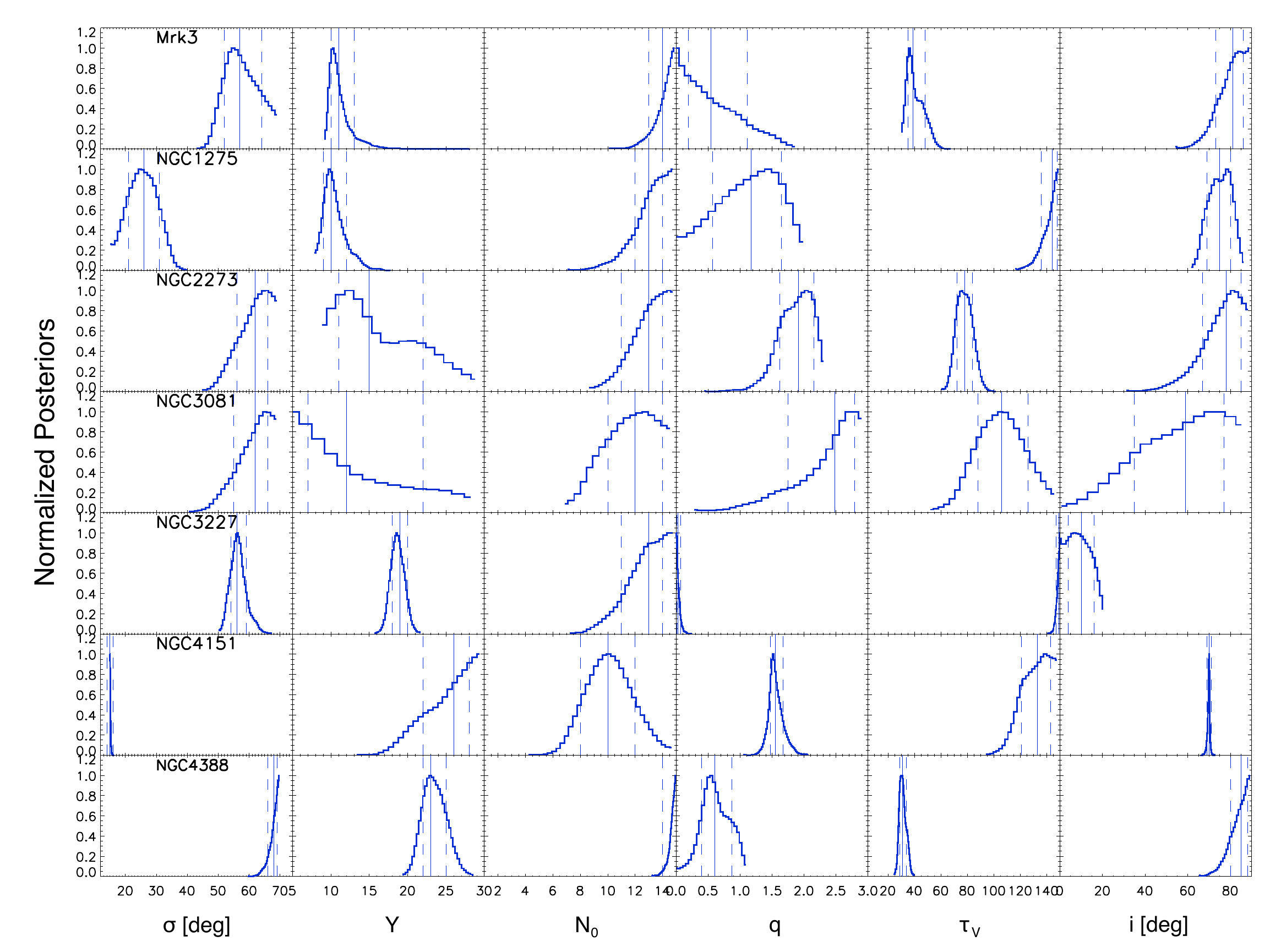}
\centering
\caption{\textsc{BayesClumpy} posterior distribution outputs.  The columns show each of the output parameters - $\sigma$, $Y$, $N_{\mbox{\tiny 0}}$, $q$, $\tau_{\mbox{\tiny V}}$, $i$ - respectively.  The solid vertical line marks the median value, while the dashed vertical lines mark the 1$\sigma$ uncertainty.}
\label{posteriors}
\end{figure*}

The posterior numerical outputs are given in Table \ref{parameters} and the distributions are shown in Figure \ref{posteriors}.  The average posterior outputs for NGC 3227, NGC 4151, and NGC 4388 tend to have narrow distributions.  For example, the average uncertainty in the outputs (excluding the error in $q$) for these AGNs is $\pm$3, while the average uncertainty in the outputs of the remaining four AGNs in the sample is $\pm$6.  These three AGNs also show a tentative turnover in torus emission between 30 - 40 $\mu$m, suggesting that sampling with these longer wavelengths produces more precise determinations of torus parameters.  However, a larger sample is needed in order to confirm this correlation.

Using the relation $R_{\mbox{\tiny out}}$ = $YR_{\mbox{\tiny in}}$, the $Y$ parameter yields torus outer radii $\sim$ 1 - 6 pc, consistent with observations \citep{Jaffe2004,Packham2005,Tristram2007,Radomski2008,GB2016}, and the estimates of previous clumpy torus model SED fittings (e.g. \citealp{RA2009,RA2011,AH2011, Ichikawa2015}).  

\subsection{Constraint of Extended Emission Source in FIR}
\label{FIR}

The contribution of dust emission at wavelengths > 30 $\mu$m is often attributed solely to star formation.  To explore the possible components contributing to FIR emission, we use available $Herschel$ data for NGC 4151 \citep{Siebenmorgen2015,GG2016} and compare it to a predicted FIR contribution from the NLR as well as star formation.  Using the torus SED output from \textsc{BayesClumpy}, we combine torus SED with 1) a scaled starburst template\footnote{\url{https://sites.google.com/site/decompir/}}, and 2) a 75 K blackbody representative of continuum NLR emission (see temperature discussion in Section \ref{dusty}).  The starburst template \citep{Mullaney2011} uses $Spitzer$/IRS spectra of an averaged subset of starburst nuclei  \citep{Brandl2006}, then extrapolates FIR emission out to 100 $\mu$m using $IRAS$ photometry.  The starburst template and blackbody are scaled to the residual 37.1 $\mu$m flux of NGC 4151, which we use here as a lower limit.  The total FIR emission is then compared to $Herschel$ 70 and 100 $\mu$m fluxes \citep{Siebenmorgen2015}, using the $Herschel$ fluxes as upper limits.

\citet{GG2016} used $Herschel$ data to identify nearby AGNs with 70 $\mu$m emission dominated by dust heated by the AGN and estimated a range of the AGN flux at 70 $\mu$m.  They estimated that 49 - 60 \% of the flux in NGC 4151 within 1kpc is due to dust heated by the AGN (both torus and NLR).  That equates to a 70 $\mu$m flux due to dust heated by the AGN at approximately 2.4 - 3.0 Jy.  

\begin{figure}
\includegraphics[scale=0.3]{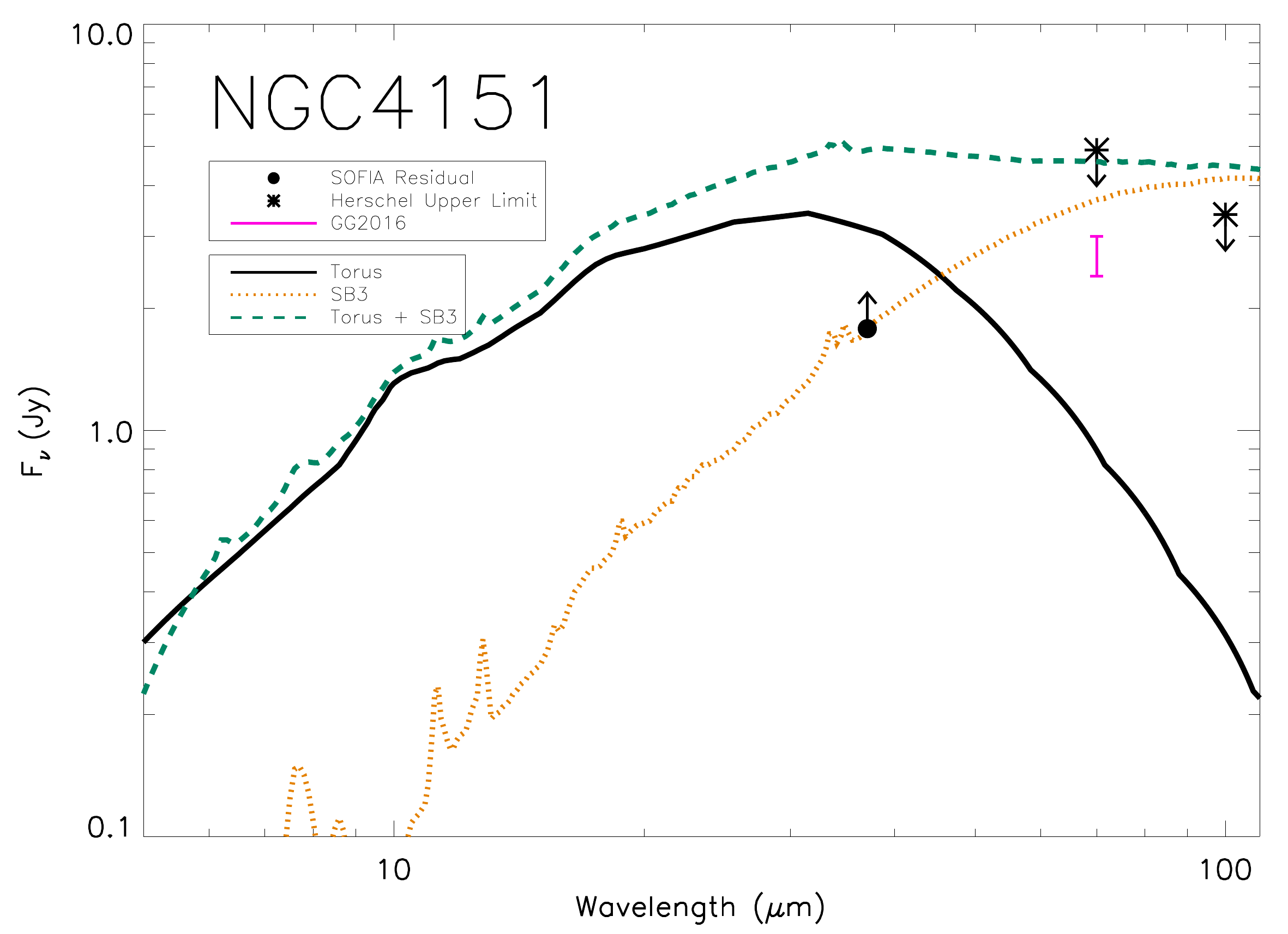}
\includegraphics[scale=0.3]{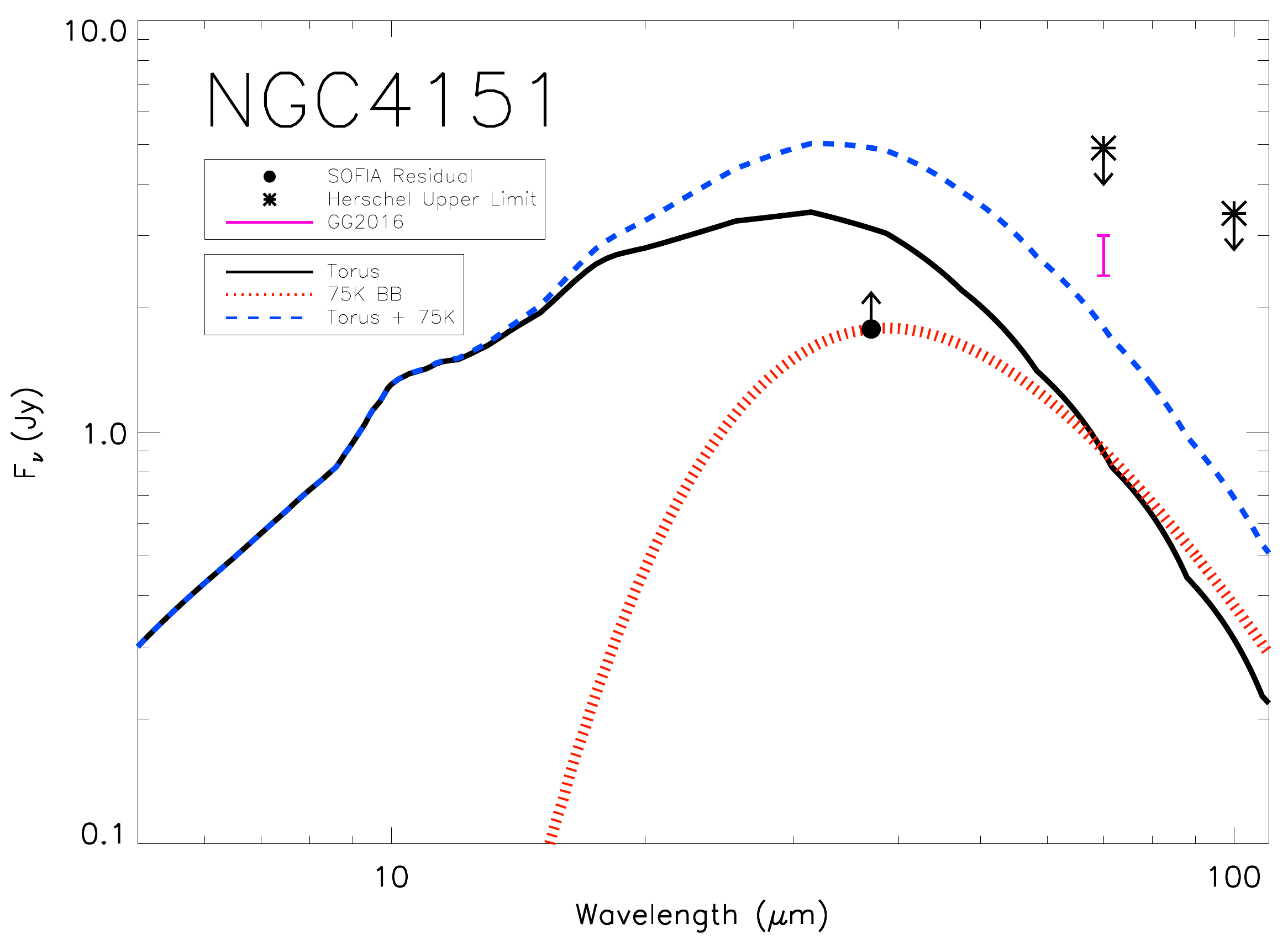}
\caption{Top: Total FIR SED representation of the torus with starburst templates SB3 \citep{Mullaney2011} scaled to the residual emission of NGC 4151 at 37.1 $\mu$m.  Bottom: Total FIR SED of the torus with a 75 K blackbody.  70 and 100 $\mu$m Herschel fluxes within a 6-7" radius are shown as an upper limit to the total FIR emission. SOFIA residual fluxes are given as a lower limit.}

\label{herschel}
\end{figure}

The top panel of Figure \ref{herschel} shows the total FIR emission when adding the starburst template SB3 (orange dotted line) to the torus SED (solid black line).  The stellar template SB3, which represents host-galaxy emission, was selected because of its relatively shallow FIR slope compared to other starburst templates.  The total emission using the Torus + SB3 (green dashed line) exceeds the 100 $\mu$m upper limit set by $Herschel$, as well as the predicted AGN contribution range of \citet{GG2016} (shown in pink).  The bottom panel of Figure \ref{herschel} shows total FIR emission when combining the 75 K blackbody (red dotted line) with the torus SED (solid black line).  The total emission of the system, Torus + 75K (blue dashed line), fits well within the upper limits set by $Herschel$ and better represents FIR emission.  The total emission also does not exceed the predicted AGN emission within 1 kpc.  

 In NGC 4151, FIR emission likely has some contribution from star formation or host galaxy which fit well within the $Herschel$ upper limits.  Assuming a single source of emission in the residuals is likely not physically accurate.  In fact, a variety of blackbody and star formation template scaling can describe the upper limit FIR emission out to 100 $\mu$m.  However, more data is needed to accurately describe the full FIR SED.  

\section{Summary and Conclusions}
\label{conclusions}

We have presented new 37.1 $\mu$m imaging observations from the SOFIA telescope for 7 AGNs.  Of these 7, 3 were also observed using the 31.5 $\mu$m filter.  To estimate torus emission within the aperture of SOFIA, we used the PSF-scaling method described by \citetalias{F16}.  Three of the 1 - 40 $\mu$m SEDs tentatively show a turnover in torus emission, suggesting that observations in the 30 - 40 $\mu$m range are needed to find peak torus emission.

We examined the origin of extended emission and show, for the first time, extended 37.1 $\mu$m emission in Mrk 3, NGC 4151, and NGC 4388 that is consistent with the NLR and radio axes.  Spectra from the $Spitzer$ CASSIS database \citep{Lebouteiller2011} generally show either strong PAH features, which is attributed to star forming regions, or a strong [O \textsc{iv}] 25.9 $\mu$m fine structure line, suggesting emission from the NLR.  We find that extended emission within the 3 - 4" FWHM of SOFIA can generally be attributed to either star formation or some elongated component consistent with the NLR.

The elongated NLR component in our observations suggests that dust can be heated by the AGN on scales of hundreds of parsecs.  In order to demonstrate this viability, we obtained a tentative estimate of the temperature of the emitting source of our residuals by fitting a blackbody function to the residual fluxes.  From the blackbody temperatures, we estimate that dust of that temperature can be heated by the AGN out to scales of hundreds of parsecs.  Using the blackbody temperatures, we also obtained dust masses from the emitting region between $\sim$ 10$^{3}$ - 10$^{5}$ M$_{\sun}$.

We compiled available NIR and MIR flux data from the literature, then used the \textsc{Clumpy} torus models of \citet{Nenkova2008a,Nenkova2008b} and the Bayesian inference tool \textsc{BayesClumpy} \citep{AR2009} to fit the IR 1 - 37 $\mu$m SEDs to obtain torus model parameters.  The posterior distributions are relatively narrow for NGC 3227, NGC 4151, and NGC 4388.  These AGN also tentatively show a turnover in torus emission, suggesting that sampling at longer wavelengths produces more precise determinations of torus parameters.  

Using the torus SED output from \textsc{BayesClumpy}, we reproduced FIR emission in NGC 4151 by adding the torus SED to two separate components.  We then compared the total emission to 70 and 100 $\mu$m $Herschel$ fluxes used as upper limits.  We first added the torus SED to a starburst template, representing star formation near the nucleus.  We then added the torus SED to a 75K blackbody, which represents dust in the NLR heated by the AGN.  We find that scaling the starburst template to the residual emission overestimates FIR emission, while scaling the blackbody representing the NLR continuum fits a more accurate description.  

\section*{Acknowledgements}

Based on observations made with the NASA/DLR Stratospheric Observatory For Infrared Astronomy (SOFIA).  SOFIA is jointly operated by the Universities Space Research Association, Inc. (USRA), under NASA contract NAS2-97001, and the Deutsches SOFIA Institut (DSI) under DLR contract 50 OK 0901 to the University of Stuttgart.  Financial support for this work was provided by NASA through award \#02\_0035, \#04\_0048, and \#06\_0066 issued by USRA.  L.F. and C.P. acknowledge support from the NSF-grant number 1616828.  CRA acknowledges the Ram\'on y Cajal Program of the Spanish Ministry of Economy and Competitiveness through project RYC-2014-15779 and the Spanish Plan Nacional de Astronom\' ia y Astrofis\' ica under grant AYA2016-76682-C3-2-P.  T.D.- S. acknowledges support from ALMA-CONICYT project 31130005 and FONDECYT regular project 1151239.  









\bsp	
\label{lastpage}
\end{document}